\documentclass{article}

\usepackage{authblk}
\usepackage[utf8]{inputenc} 
\usepackage[T1]{fontenc}    
\usepackage{hyperref}       
\usepackage{url}            
\usepackage{booktabs}       
\usepackage{amsfonts}       
\usepackage{nicefrac}       
\usepackage{microtype}      
\usepackage{lipsum}		
\usepackage{graphicx}
\usepackage{natbib}
\usepackage{doi}
\usepackage{amsmath, amsthm}
\usepackage{multirow}
\usepackage{multicol}
\usepackage{placeins}
\usepackage{enumitem}

\usepackage[verbose=true,letterpaper]{geometry}
\AtBeginDocument{
  \newgeometry{
    textheight=9in,
    textwidth=6.5in,
    top=1in,
    headheight=14pt,
    headsep=25pt,
    footskip=30pt
  }
}

\usepackage{setspace}
\newtheorem*{theorem*}{Theorem}

\newcommand{\robust}{\mbox{\tiny{robust}}}

\title{Demystify Doubly-Robust Estimation: The Role of Overlap}

\author[1]{Chengxin Yang}
\author[1,2]{Laine E. Thomas}
\author[3]{Fan Li}
\affil[1]{Department of Biostatistics and Bioinformatics, Duke University, Durham NC 27710}
\affil[2]{Duke Clinical Research Institute, Durham NC 27701}
\affil[3]{Department of Statistical Science, Duke University, Durham NC 27708}

\begin{document}
\maketitle

\begin{abstract}
\textbf{Objectives:} The doubly-robust (DR) estimator is popular for evaluating causal effects in observational studies and is often perceived as more desirable than inverse probability weighting (IPW) or outcome modeling alone because it provides extra protection against model misspecification. However, double robustness is an asymptotic property that may not hold in finite samples. We investigate how the finite sample performance of the DR estimator depends on the degree of covariate overlap between comparison groups.

\textbf{Methods:} Using analytical illustrations and extensive simulations under various scenarios with different degrees of covariate overlap and model specifications, we examine the bias and variance of the DR estimator relative to IPW and outcome modeling estimators.

\textbf{Results:} We find that: (i) specification of the outcome model has a stronger influence on the DR estimates than specification of the propensity score model, and this dominance increases as overlap decreases; (ii) with poor overlap, the DR estimator generally amplifies the adverse consequences of extreme weights (large bias and/or variance) regardless of model specifications, and is often inferior to both the IPW and outcome modeling estimators.

\textbf{Conclusions:} As a practical guide, we recommend always first checking the degree of overlap in applications. In the case of poor overlap, analysts should consider shifting the target population to a subpopulation with adequate overlap via methods such as trimming or overlap weighting.
\end{abstract}


\section{Introduction} 

In causal inference, there are two general approaches for estimating average treatment effects: outcome modeling and inverse probability weighting (IPW). A third approach, the doubly-robust (DR) estimation \cite{robins1999dropout, robins1994regcoef, L&D2004compareweights, Bang&Robins2005DoubleRobust}, combines the two methods by augmenting an IPW estimator with an outcome model or vice versa, and is also known as the augmented IPW (AIPW) estimator. The DR estimator is consistent, that is, asymptotically unbiased with bounded variance, for the average treatment effect, if either the propensity score model or the outcome model, but not necessarily both, is correct. Therefore, the DR estimator is usually perceived as more desirable because it provides extra protection against model misspecification.

However, double robustness is an asymptotic property that may not hold in finite samples. In particular, the augmentation in the DR estimator does not address a well-known weakness of the IPW estimator, namely, the sensitivity to propensities close to 0 or 1, which leads to extreme weights and, consequently, large bias and variance in finite samples. Despite the large literature on the DR estimation, the finite sample performance of the DR estimator has been rarely examined. In this paper, we scrutinize the perception of the DR estimator being superior in the context of finite sample performance. As we shall show, the DR estimator not only inherits but also exacerbates the weakness due to extreme propensities when covariates between the comparison groups are poorly overlapped. More generally, the finite sample performance of the DR estimator critically depends on the degree of covariate overlap. In cases of limited overlap, a shift to methods focusing on target populations with sufficient overlap, such as trimming and overlap weighting, usually yields more robust, efficient, and precise causal estimates.

\section{Methods}

\subsection{Notations, estimands, and estimators}

Consider an observational study with $N$ subjects. For each subject $i(= 1, \ldots, N)$, we observe a set of pre-treatment covariates $X_i$, treatment status $Z_i(= z)$, with $z = 1$ for treatment and $z = 0$ for control, and an outcome $Y_i$. Assuming no interference and consistency \cite{Rubin1980RerandFisher}, or the stable unit treatment value assumption (SUTVA), each subject has two potential outcomes $Y_i(1)$ and $Y_i(0)$, corresponding to treatment and control, respectively, one of which is observed $Y_i = Y_i(Z_i)$. Throughout, we assume the existence of $\{Y_i(1), Y_i(0)\}$, and thus the individual treatment effect $Y_i(1) - Y_i(0)$. The propensity score (PS) is the probability of receiving treatment given covariates: $e(X) = P(Z = 1|X)$ \cite{Rosenbaum&Rubin1983propensity}. The PS is unknown in observational studies and must be estimated, e.g., via a logistic regression $e(X, \beta) = 1/(1 + \exp(-\beta^T X))$, where $\beta$ is the model parameter. The standard target estimand is the average treatment effect (ATE): $\tau = E(Y_i(1) - Y_i(0))$, where the expectation is over the target population represented by the study sample. Throughout, we maintain two standard assumptions in causal inference: i) positivity, i.e., each unit has a non-zero probability of being assigned to either treatment, ii) conditional exchangeability, i.e., there is no unmeasured confounder \cite{hernan2023causalbook}. Under these assumptions, we have two identification formulas for $\tau$:
\begin{equation}
\tau = E[E(Y_i|Z_i = 1, X_i) - E(Y_i|Z_i = 0, X_i)] = E \left[\frac{Y_i Z_i}{e(X_i)} - \frac{Y_i(1 - Z_i)}{1 - e(X_i)}\right].
\label{eq:identification}
\end{equation}

Accordingly, there are two types of estimators. The first is the outcome-model estimator, where we specify a model for the potential outcome, $\mu_z(x) = E(Y_i(z)|X_i = x) = E(Y_i|Z_i = z, X_i = x)$, and then estimate $\tau$ by $\hat{\tau}_{om} = \sum_{i=1}^{N}(\hat{\mu}_1(X_i) - \hat{\mu}_0(X_i))/N$, where $\hat{\mu}_z(X_i)$ is the fitted outcome for treatment $z$ under the outcome model. The second is the IPW estimator,
\begin{equation*}
\hat{\tau}_{ipw} = \frac{1}{N} \left\{\sum_{i=1}^{N} \frac{Z_i Y_i}{\hat{e}(X_i)} - \sum_{i=1}^{N} \frac{(1 - Z_i) Y_i}{1 - \hat{e}(X_i)}\right\},
\end{equation*}
where $\hat{e}(X)$ is the model-based estimated propensity score, and the weight is $1/\hat{e}(X_i)$ for treated subjects and $1/(1 - \hat{e}(X_i))$ for control subjects. Another version of the IPW estimator, called the Hajek estimator, normalizes the weights so that the sum of all weights in one group is 1. Both $\hat{\tau}_{om}$ and $\hat{\tau}_{ipw}$ require the corresponding model to be correctly specified for consistency for $\tau$.

The doubly-robust (DR) estimator is:
\begin{equation}
\hat{\tau}_{dr} = \frac{1}{N}\sum_{i=1}^{N}\left\{\hat{\mu}_1(X_i) + \frac{Z_i(Y_i - \hat{\mu}_1(X_i))}{\hat{e}(X_i)} \right\} - \frac{1}{N}\sum_{i=1}^{N}\left\{\hat{\mu}_0(X_i) + \frac{(1 - Z_i)(Y_i - \hat{\mu}_0(X_i))}{1 - \hat{e}(X_i)} \right\}.
\label{eq:dr}
\end{equation}
One can prove that $E(\hat{\tau}_{dr}) = \tau$ if either the outcome model or the PS model, but not necessarily both, is correctly specified \cite{Bang&Robins2005DoubleRobust}, and thus $\hat{\tau}_{dr}$ provides two chances of ``getting it right.''

\subsection{Finite-sample error of the doubly-robust estimator}

The finite-sample inference treats the treatment assignment $Z_i$ as the only source of randomness while regards $\{Y_i(1), Y_i(0), X_i\}_{i=1}^{N}$, the study samples at hand, as non-random. Accordingly, the sample ATE (SATE), $\tau_S = N^{-1} \sum_{i=1}^{N} \{Y_i(1) - Y_i(0)\}$, is the standard finite-sample estimand. In contrast, the large-sample framework treats study samples as random draws from a super-population and uses $\tau$, the average treatment effect over this super-population (PATE), as the estimand in place of its finite-sample counterpart $\tau_S$. Specifically, the large-sample framework consists of two layers of randomness \cite{Miratrix2018WeightSurvey, Abadie2020samplingVSdesign, PD2024causalbook}: i) the sampling process of the study subjects, which accounts for randomness in $X_i$ and $\tau - \tau_S$; ii) random assignment of the treatments. In contrast, the finite-sample framework only consists of the latter.

Throughout, we maintain $\tau$, the population average treatment effect, as our target estimand, in consistency with the large-sample framework in which DR is theoretically rooted and the vast literature of observational studies where DR has been commonly applied. Nevertheless, it's still of great value to set aside the difference between $\tau$ and $\tau_S$ due to the random sampling and investigate the performance of DR in approximating $\tau_S$ from a finite-sample perspective. As we will introduce soon, it provides deep insights into the finite-sample behaviors of DR and their implications for large-sample inference. Moreover, being a good approximation to $\tau_S$ is an intrinsic requirement of common scientific induction: the statistical analysis results should first be \emph{internally valid} for the study samples at hand before being generalized to any broader population beyond the study samples \cite{PD2024causalbook}.

Ideally, a good estimator should yield estimate close to $\tau_S$. The difference between the DR estimator and SATE, $\Delta_{dr} = \hat{\tau}_{dr} - \tau_S$, is:
\begin{equation}
\Delta_{dr} := \frac{1}{N}\sum_{i=1}^{N} \frac{Z_i - \hat{e}(X_i)}{\hat{e}(X_i)} \{Y_i(1) - \hat{\mu}_1(X_i)\} + \frac{1}{N}\sum_{i=1}^{N} \frac{Z_i - \hat{e}(X_i)}{1 - \hat{e}(X_i)} \{Y_i(0) - \hat{\mu}_0(X_i)\}.
\label{eq:delta}
\end{equation}
We term $\Delta_{dr}$ the \emph{finite-sample error of the DR estimator} to indicate that it's the error relative to $\tau_S$. $\Delta_{dr}$ is model-based and can be decomposed as: $\Delta_{dr} = \frac{1}{N} \sum_i^{N} R_i^e R_i^y$, where
\begin{equation*}
R_i^e = Z_i - \hat{e}(X_i) \quad \text{and} \quad R_i^y = \frac{Y_i(1)-\hat{\mu}_1(X_i)}{\hat{e}(X_i)} + \frac{Y_i(0)-\hat{\mu}_0(X_i)}{1-\hat{e}(X_i)},
\end{equation*}
are the residuals of the PS model and the weighted residuals of the outcome model for unit $i$, respectively, and we refer to them as \emph{residuals} henceforth. It's worth clarifying that none of $\tau_S$, $R_i^y$, or $\Delta_{dr}$ is observable, due to the missingness of the counterfactual outcomes. However, this viewpoint and the decomposition of $\Delta_{dr}$ elucidate the finite-sample behaviors of the DR estimator and corresponding large-sample implications, and moreover, enable us to demonstrate and verify these intuitions through simulations, in which we generate and keep track of all the counterfactual outcomes.

Importantly, the above decomposition elucidates the critical role of overlap in the DR estimation. Specifically, the PS residual $R_i^e$ exists for each unit even with a correctly specified PS model, and is particularly large for units whose actual treatment status is opposite to what the PS predicts. The outcome residual $R_i^y$ for each unit is the sum of the prediction error of the potential outcome under each treatment level weighted by the inverse of the estimated probability of that treatment. Therefore, in finite samples, $R_i^y$ will be much amplified for units with the estimated PS close to 0 or 1, i.e., regions with little overlap between treatment levels. Though such \emph{error amplification} by extreme propensities exists regardless of whether the outcome model is correctly specified, it is particularly severe when the outcome model is misspecified, because the prediction error due to extrapolation by a misspecified outcome model in regions with little overlap can be excessive \cite{zhou2020PSmisspec}. The above discussion shows the double jeopardy to the DR estimation exerted by the lack of overlap.

In summary, though the DR estimator provides additional protection against model misspecification in theory, it also inherits and exacerbates the finite-sample adverse consequences of insufficient overlap in both IPW and outcome modeling strategies. Therefore, the finite-sample error of the DR estimator cannot be reduced by merely specifying the PS model correctly; it remains crucial to correctly specify the outcome model, which is a challenging task. Arguably, if a correct outcome model is available, one would prefer the corresponding outcome model estimator over its augmented DR version because the former is consistent and asymptotically more efficient than the latter \cite{Tan2007CommentsDR, Robins2007CommentsKS}, and thus defeats the purpose of a DR estimator.

To mitigate the adverse impact of lack of overlap in an observational study, the literature has shifted to alternative methods that focus on target populations (and correspondingly, estimands) with adequate covariate overlap. This can be achieved using, e.g., the DR estimator with trimmed weights \cite{Crump2009Trimming}, overlap weighting \cite{Li2018overlap, Li2018ExtremePS}, or matching estimators \cite{LiGreene2013weightanalogue, Yoshida2017Matchingweights}.

\section{Simulation studies}

\subsection{Simulation design}

We conduct simulation studies under various scenarios with different degrees of covariate overlap between groups. Following the simulation design in Li et al. \cite{Li2018ExtremePS} , we first generate $V_1$--$V_6$ from a multivariate normal distribution with zero mean, unit marginal variance, and pairwise correlation of 0.5. Then, we retain $V_1$--$V_3$ as $X_1$--$X_3$, while dichotomizing $V_4$--$V_6$ by indicator function $I_{\{V<0\}}$ to create binary $X_4$--$X_6$, with each having a marginal expectation of approximately 0.5. The true propensity score follows a logistic model, $\text{logit}(e(X)) = \alpha_0 + \alpha_1 X_1 + \alpha_2 X_2 + \alpha_3 X_3 + \alpha_4 X_4 + \alpha_5 X_5 + \alpha_6 X_6$, and the observed binary treatment is generated independently from a Bernoulli distribution with success probability $e(X)$. The continuous outcome $Y$ follows $Y = \beta_0 + \beta_1 X_1 + \beta_2 X_2 + \beta_3 X_3 + \beta_4 X_4 + \beta_5 X_5 + \beta_6 X_6 + \tau Z + \epsilon$, with $\epsilon$ being a standardized normal distribution, implying a homogeneous treatment effect, $\tau$, for all subjects. For simplicity, we set $\tau = 1$.

We simulate two treatment prevalences and two degrees of overlap, constituting four scenarios. Specifically, we set the PS model parameters as $(\alpha_1, \alpha_2, \alpha_3, \alpha_4, \alpha_5, \alpha_6) = d \times (0.2, 0.3, 0.4, -0.25, -0.3, -0.3)$. We choose the intercept $\alpha_0$ for treatment prevalence of approximately 0.4 and 0.1, and $d$ to be 1 and 3 for good and poor covariate overlap, respectively. More descriptions of the distribution of PS for each scenario are provided in the Supplementary Material. For the outcome model, we choose $(\beta_1, \beta_2, \beta_3, \beta_4, \beta_5, \beta_6) = (-0.5, -0.8, -1.2, 0.8, 0.8, 1)$ and $\beta_0 = 0$, so that subjects with lower outcomes are more likely to receive the treatment. We repeatedly simulate 10,000 datasets of a range of sample sizes $n \in \{100, 300, 500, 1000, 2000\}$ under each scenario, and report our primary results based on $n = 500$. The sample size of 100 is not applied to the treatment prevalence of 0.1 due to a non-negligible proportion of sampled datasets containing no treated individuals. For each scenario, we estimate $\tau$ by IPW and DR, without and with trimming of PS at 0.05 or 0.1, as well as overlap weighting and outcome regression. All postulated outcome models are linear. The number of simulation replicates, 10,000, is generally large enough for numerically stable results for our primary results; the only exception is that, under poor overlap, DR without trimming generates results with moderate numerical difference under different random seeds, but our observed phenomena and patterns remain unchanged.

We further consider model misspecification. We simulate two types of misspecifications, i) wrong functional form (mistaking $X_3$ by $X_3^2$) and ii) omitting variables ($X_3$ and $X_4$). Each type of misspecification is applied within either the propensity or outcome model, but not both, resulting in four model misspecifications. We report the primary results based on the wrong functional form and the Supplementary Material provides results under omitting variables. In general, wrong functional form refers to any situation when a covariate is included in the regression model in a form different from the correct form in which it factors into the covariate-outcome relationship. We use the wrong polynomial term as one possibility only for the sake of illustration.

\subsection{Simulation results}

We arrange our presentations of the simulations as follows: Tables~\ref{tab:Allest-compare} and \ref{tab:Misspec-wrongfuncform} display the results using various estimators in estimating $\tau$. Table~\ref{tab:FSerror-wrongfuncform} and all figures afterward display finite-sample quantities to provide insights into Tables~\ref{tab:Allest-compare} and \ref{tab:Misspec-wrongfuncform} and illustrate the finite-sample behaviors of the DR estimator. We refer to the estimator without trimming when mentioning its name solely.

\begin{table}[!ht]
\centering
\caption{Root mean square error (RMSE) and bias for estimators with correct model specifications under various degrees of overlap.}
\begin{tabular}{lcccccccc}
\midrule
 \multirow{3}{*}{Estimator} & \multicolumn{4}{c}{Treatment prevalence = 0.4} & \multicolumn{4}{c}{Treatment prevalence = 0.1} \\
 & \multicolumn{2}{c}{$d = 1$} & \multicolumn{2}{c}{$d = 3$} & \multicolumn{2}{c}{$d = 1$} & \multicolumn{2}{c}{$d = 3$} \\ \cmidrule{2-9}
 & RMSE & BIAS & RMSE & BIAS & RMSE & BIAS & RMSE & BIAS \\ \midrule
Doubly-robust &  &  &  &  &  &  &  &  \\
\quad No trimming & 0.11 & 0.00 & 0.47 & 0.00 & 0.24 & 0.00 & 2.31 & $-$0.01 \\
\quad $\delta = 0.05$ & 0.11 & 0.00 & 0.16 & 0.00 & 0.20 & 0.00 & 0.25 & $-$0.01 \\
\quad $\delta = 0.1$ & 0.11 & 0.00 & 0.16 & 0.00 & 0.22 & 0.00 & 0.25 & $-$0.01 \\
IPW &  &  &  &  &  &  &  &  \\
\quad No trimming & 0.19 & 0.00 & 1.04 & $-$0.37 & 0.65 & $-$0.05 & 2.00 & $-$1.52 \\
\quad $\delta = 0.05$ & 0.16 & 0.00 & 0.20 & $-$0.01 & 0.26 & $-$0.01 & 0.29 & $-$0.03 \\
\quad $\delta = 0.1$ & 0.13 & 0.00 & 0.17 & 0.00 & 0.25 & $-$0.01 & 0.26 & $-$0.02 \\
Outcome regression & 0.10 & 0.00 & 0.14 & 0.00 & 0.21 & 0.00 & 0.51 & $-$0.02 \\
Overlap weighting & 0.10 & 0.00 & 0.14 & 0.00 & 0.16 & 0.00 & 0.20 & 0.00 \\ \midrule
\end{tabular}
\label{tab:Allest-compare}
\end{table}

Table~\ref{tab:Allest-compare} displays the root mean squared error (RMSE) and bias of each estimator when both the outcome and propensity model are correctly specified. Under an approximately balanced design (prevalence 0.4) and sufficient overlap ($d = 1$), all estimators have negligible bias, with the outcome regression and overlap weighting estimators having the smallest RMSE. As the overlap deteriorates ($d = 3$), IPW becomes biased, and IPW and DR lead to a sharp increase in RMSE, whereas other estimators remain generally unbiased and stable. Notably, though DR generally improves over the IPW, it is universally worse in all scenarios than the outcome regression, as well as the methods focusing on the overlapped population, namely, its counterpart with trimming and overlap weighting.

\begin{table}[!ht]
\caption{RMSE and bias for estimators with one of the propensity and outcome models being misspecified, or both models being misspecified. The misspecification is incurred by wrong functional form (mistake $X_3$ by $X_3^2$).}
\resizebox{\textwidth}{!}{
\begin{tabular}{lllcccccccc}
\midrule
 \multicolumn{2}{c}{\multirow{3}{*}{Estimator}} & \multirow{3}{*}{\shortstack{Misspec.\\model}} & \multicolumn{4}{c}{Treatment prevalence = 0.4} & \multicolumn{4}{c}{Treatment prevalence = 0.1} \\
 \multicolumn{2}{c}{} &  & \multicolumn{2}{c}{$d = 1$} & \multicolumn{2}{c}{$d = 3$} & \multicolumn{2}{c}{$d = 1$} & \multicolumn{2}{c}{$d = 3$} \\ \cmidrule{4-11}
 \multicolumn{2}{c}{} &  & RMSE & Bias & RMSE & Bias & RMSE & Bias & RMSE & Bias \\ \midrule
\multirow{6}{*}{\shortstack{Doubly-\\robust}} & \multirow{3}{*}{No trimming} & PS & 0.11 & 0.00 & 0.24 & 0.00 & 0.23 & 0.00 & 0.89 & $-$0.01 \\
 &  & Outcome & 0.14 & $-$0.02 & 2.11 & $-$0.07 & 0.43 & $-$0.08 & 4.06 & $-$0.45 \\
 &  & Both & 0.36 & $-$0.33 & 1.07 & $-$0.98 & 0.46 & $-$0.33 & 1.48 & $-$0.78 \\ \cmidrule{2-11}
 & \multirow{3}{*}{$\delta = 0.05$} & PS & 0.11 & 0.00 & 0.16 & 0.00 & 0.20 & 0.00 & 0.27 & $-$0.01 \\
 &  & Outcome & 0.13 & $-$0.01 & 0.19 & $-$0.02 & 0.23 & $-$0.02 & 0.26 & $-$0.01 \\
 &  & Both & 0.36 & $-$0.33 & 0.86 & $-$0.84 & 0.36 & $-$0.25 & 0.47 & $-$0.38 \\ \midrule
\multirow{2}{*}{IPW} & No trimming & PS & 0.35 & $-$0.30 & 1.22 & $-$1.06 & 0.60 & $-$0.32 & 2.03 & $-$1.65 \\
 & $\delta = 0.05$ & PS & 0.34 & $-$0.30 & 0.81 & $-$0.78 & 0.38 & $-$0.24 & 0.50 & $-$0.38 \\ \midrule
\multicolumn{2}{l}{Outcome regression} & Outcome & 0.40 & $-$0.37 & 1.09 & $-$1.08 & 0.47 & $-$0.36 & 1.13 & $-$0.99 \\ \midrule
\multicolumn{2}{l}{Overlap weighting} & PS & 0.31 & $-$0.28 & 0.76 & $-$0.74 & 0.31 & $-$0.23 & 0.44 & $-$0.37 \\
\midrule
\end{tabular}
}
\noindent{\footnotesize Results from trimming with $\delta = 0.1$ are similar to that with $\delta = 0.05$ and are thus not displayed.}
\label{tab:Misspec-wrongfuncform}
\end{table}

Table~\ref{tab:Misspec-wrongfuncform} displays the RMSE and bias of each estimator under misspecification of wrong functional form. The patterns under misspecification of omitting variables are similar and are thus delegated to the Supplementary Material. Despite that, theoretically, the DR estimator is consistent for $\tau$ under all these situations, the finite-sample performance varies significantly. With sufficient overlap and balanced design, correctly specifying either the PS or outcome model leads to negligible bias. On the contrary, with poor overlap and/or an imbalanced design, the outcome model specification exerts a much stronger influence than the PS model specification, and the dominance increases as the degree of overlap decreases. Specifically, compared to DR with both models being correct (Table~\ref{tab:Allest-compare}), DR with only a correct outcome model yields similar results, whereas DR with only a correct propensity model yields a considerable increase in bias and, particularly, RMSE, which can be even manyfold as the estimand under poor overlap. This illustrates that poor overlap dramatically diminishes the extra protection provided by a correct propensity model in the DR estimator.

As predicted by the theory, the DR estimator with misspecification in only one of the two models largely outperforms the corresponding misspecified IPW and outcome regression estimators in most cases. However, exceptions are under severe lack of overlap ($d = 3$), where a DR estimator with misspecified outcome model renders a much larger RMSE than all other estimators. In contrast, the DR estimator with trimmed weights consistently yields small bias and variance.

Interestingly, under poor overlap, the DR estimator with both propensity and outcome models being misspecified generally yields large bias, but surprisingly has smaller RMSE than its counterpart with only the outcome model being misspecified. Moreover, when both models are misspecified, trimming of extreme weights does not improve the performance of DR as much as it does when only one model is misspecified. In summary, there is a clear interplay between overlap and model misspecification in the finite-sample performance of the DR estimator, echoing the findings in \cite{Miratrix2018WeightSurvey}.

\begin{table}[!ht]
\caption{Mean and mean absolute value (MAV) of finite-sample error in different (sub)populations with both models being correct, or one of the propensity and outcome models being misspecified by wrong functional form (mistaking $X_3$ by $X_3^2$).}
\begin{tabular}{lllcccccccc}
\midrule
 \multicolumn{2}{c}{\multirow{3}{*}{\shortstack{Model\\specification}}} & \multirow{3}{*}{Population} & \multicolumn{4}{c}{Treatment prevalence = 0.4} & \multicolumn{4}{c}{Treatment prevalence = 0.1} \\
 &  &  & \multicolumn{2}{c}{$d = 1$} & \multicolumn{2}{c}{$d = 3$} & \multicolumn{2}{c}{$d = 1$} & \multicolumn{2}{c}{$d = 3$} \\ \cmidrule{4-11}
 &  &  & Mean & MAV & Mean & MAV & Mean & MAV & Mean & MAV \\ \midrule
\multicolumn{2}{c}{\multirow{3}{*}{Correct specification}} & Overall & 0.00 & 0.07 & 0.00 & 0.22 & 0.00 & 0.18 & $-$0.01 & 0.56 \\
 &  & Tail & 0.00 & 0.32 & 0.01 & 0.87 & 0.01 & 0.51 & 0.01 & 1.02 \\
 &  & Bulk & 0.00 & 0.08 & 0.00 & 0.12 & 0.00 & 0.20 & $-$0.01 & 0.54 \\ \midrule
\multirow{6}{*}{\shortstack{Mis-\\specification}} & \multirow{3}{*}{PS model} & Overall & 0.00 & 0.07 & 0.00 & 0.16 & 0.00 & 0.17 & $-$0.01 & 0.49 \\
 &  & Tail & 0.00 & 0.29 & $-$0.02 & 0.65 & 0.00 & 0.47 & 0.02 & 0.72 \\
 &  & Bulk & 0.00 & 0.08 & 0.00 & 0.11 & 0.00 & 0.20 & $-$0.01 & 0.50 \\ \cmidrule{2-11}
 & \multirow{3}{*}{\shortstack{Outcome\\model}} & Overall & $-$0.02 & 0.10 & $-$0.07 & 0.67 & $-$0.08 & 0.31 & $-$0.45 & 1.07 \\
 &  & Tail & $-$0.14 & 0.85 & $-$0.47 & 5.08 & $-$0.42 & 1.79 & $-$0.98 & 3.02 \\
 &  & Bulk & 0.00 & 0.10 & $-$0.01 & 0.16 & $-$0.02 & 0.26 & $-$0.25 & 0.80 \\ \midrule
\end{tabular}
\label{tab:FSerror-wrongfuncform}
\end{table}

To further examine the finite-sample behavior of the DR estimator, we evaluate the finite-sample error of three (sub)populations: i) overall, which includes all subjects, ii) tail, which includes subjects with $\hat{e}(X_i)$ outside of 0.05--0.95 percentiles, and iii) bulk, which includes subjects with $\hat{e}(X_i)$ within 0.25--0.75 percentiles, corresponding to the total, poorly-overlapped, and overlapped population, respectively, as indicated by the postulated propensity score model. For each population labeled by $l$, we calculate the mean and mean absolute value (MAV) over 10,000 replications of the finite-sample error specific to (sub)population $l$, $\Delta_{dr,l}= \sum_{i:e(X_i)\in l} R_i^e R_i^y /N_l$, presented in Table~\ref{tab:FSerror-wrongfuncform}.

Based on Table~\ref{tab:FSerror-wrongfuncform}, across all model specifications and simulation scenarios, the tail population is higher (or no less) in MAV and mean of $\Delta_{dr,l}$ than the bulk population, while the overall population takes values of $\Delta_{dr,l}$ in between. Notably, a misspecified outcome model leads to an extremely sharp increase in both the mean and MAV of $\Delta_{dr,l}$ in the tail, and the severity increases as the overlap deteriorates and the design becomes imbalanced (prevalence 0.1); in contrast, $\Delta_{dr,l}$ in the bulk remains low and stable under outcome model misspecification. On the other hand, misspecification of the PS model doesn't lead to noticeable change in $\Delta_{dr,l}$ for all (sub)populations. In summary, the tail population is the primary determinant of $\Delta_{dr}$ and is extremely vulnerable to outcome model misspecification, echoing the double jeopardy of DR we discussed in the Methods Section.

\begin{figure}[!ht]
\includegraphics[width=\textwidth]{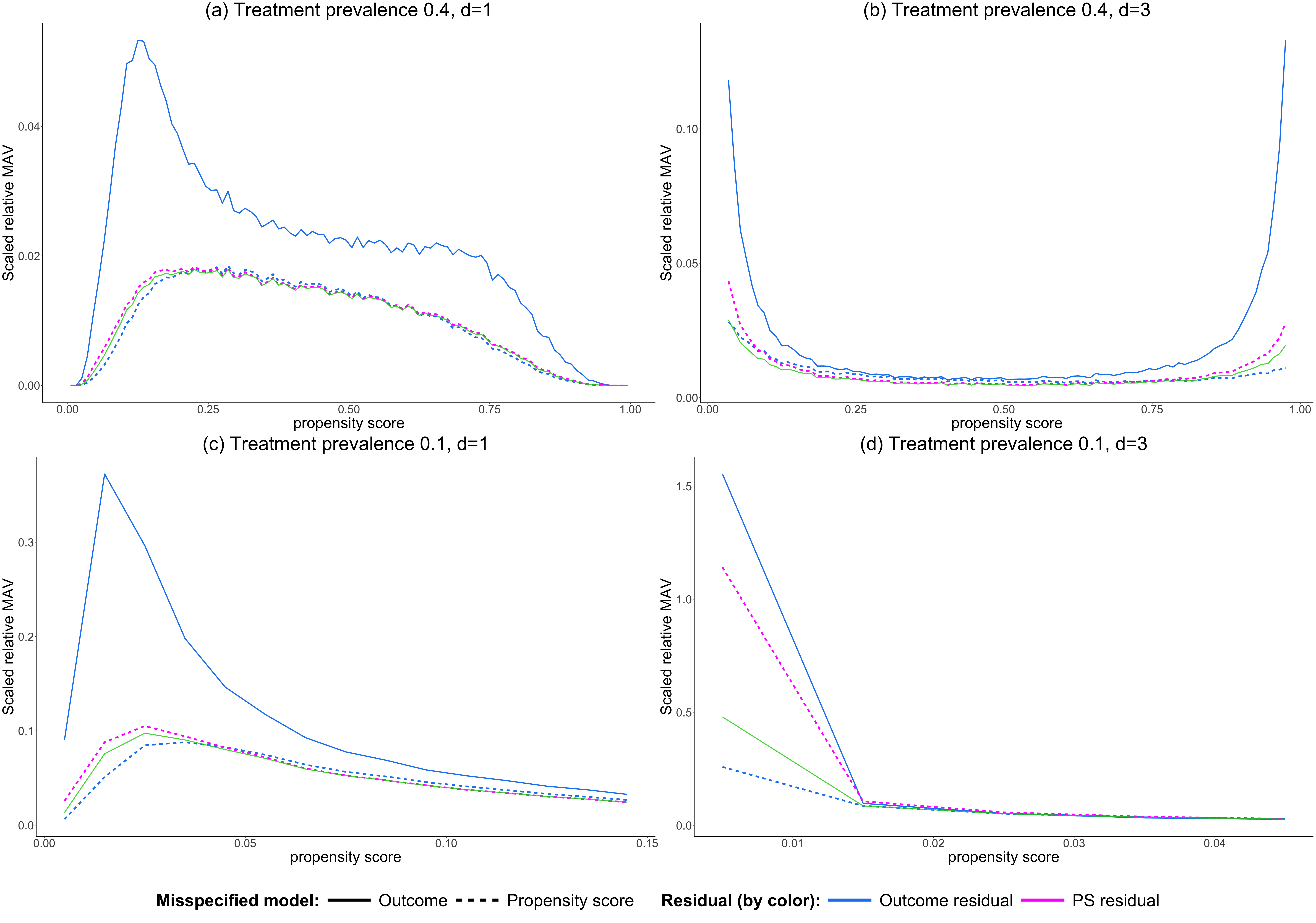}
\caption{Scaled relative MAV of $R_i^e$ and $R_i^y$. Misspecification is incurred by wrong functional form (mistaking $X_3$ by $X_3^2$). The green line ($\Phi(l)$) indicates no change in relative scale. Large values at tails of panel (b) and the rest of panels (c) and (d) are delegated to the Supplementary Material.}
\label{fig:figure1}
\end{figure}

\begin{figure}[!ht]
\includegraphics[width=\textwidth]{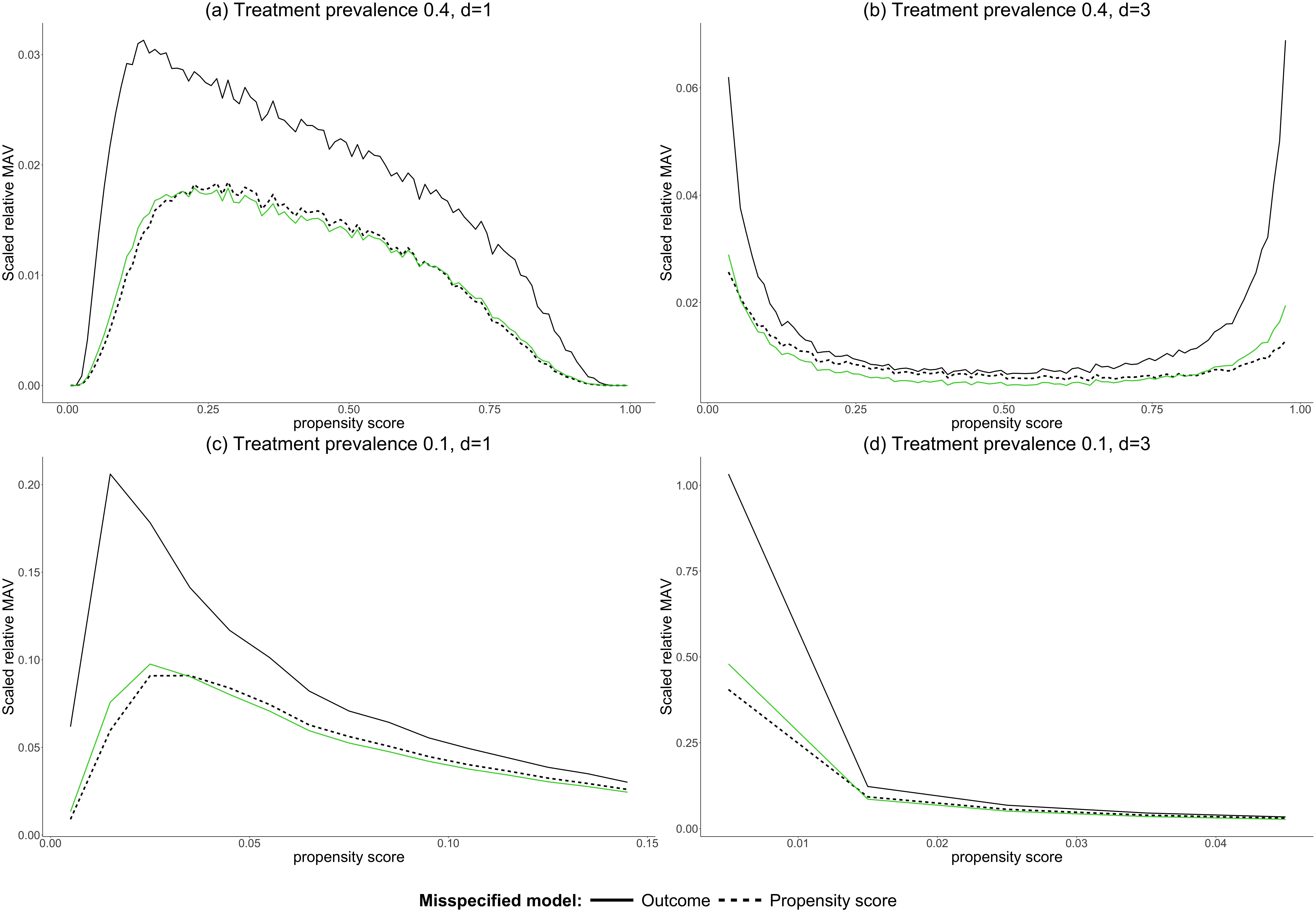}
\caption{Scaled relative MAV of $R_i^e R_i^y$. Misspecification is incurred by wrong functional form (mistaking $X_3$ by $X_3^2$). The green line ($\Phi(l)$) indicates no change in relative scale. Large values at tails of panel (b) and the rest of panels (c) and (d) are delegated to the Supplementary Material.}
\label{fig:figure2}
\end{figure}

Figures~\ref{fig:figure1} and \ref{fig:figure2} visualize the behaviors of residuals, $R_i^e$ and $R_i^y$, and their product, $R_i^e R_i^y$, across the range of propensity score, and thus provide deeper insights into model misspecification and its interplay with the degree of overlap. Specifically, we divide the true propensity score evenly into 100 intervals with a length of 0.01. For each interval $l$, we calculate the mean absolute value (MAV) over 10,000 replications of the average residuals, $\sum_{i:e(X_i)\in l} R_i^e /N_l$ and $\sum_{i:e(X_i)\in l} R_i^y /N_l$, and the interval-specific finite-sample error $\sum_{i:e(X_i)\in l} R_i^e R_i^y /N_l$, under each model specification. Then, we calculate the relative MAVs by taking the ratio of MAVs under each model misspecification to that under the correct model specification. Thus, a relative MAV larger (smaller) than 1 evaluates the extent to which the expected absolute value of $R_i^e$, $R_i^y$, or $R_i^e R_i^y$, would be inflated (deflated) in interval $l$ if that model misspecification occurs. To reflect the impact of each interval $l$ on the overall estimation, we further scale the relative MAV in each interval $l$ by multiplying $\Phi(l)$, where $\Phi(l)$ is the probability mass of the true propensity score being in interval $l$ and is estimated empirically using a large-sample dataset from the target population with a size of 100,000. The scaled relative MAV of residuals and their products are displayed in Figures~\ref{fig:figure1} and \ref{fig:figure2}, respectively; misspecification in outcome model does not change $R_i^e$ and is thus not displayed.

From Figure~\ref{fig:figure1}, model misspecifications generally impact $\Delta_{dr}$ through PS regions where the population is concentrated, as indicated by $\Phi(l)$. Among all model misspecifications and residuals, the relative increase in outcome residual caused by misspecification in the outcome model is the sharpest in all scenarios, and the extent increases as the overlap deteriorates. More specifically, with a balanced design, outcome model misspecification leads to a much sharper increase in the outcome residual at the tails of PS under insufficient overlap than that at the bulk of PS under sufficient overlap; similar comparisons could be made under an imbalanced design. These observations confirm the error amplification for individuals with PS close to 0 and 1 and clarify the consequent dominance of outcome model specification under poor overlap.

Interestingly but also notably, misspecification in the PS model does not always inflate the outcome residual -- exceptions are at subpopulations with propensities close to 0 and 1, because a misspecified propensity model does not necessarily yield inverse probability weights that are more extreme than the correct propensity model does, and thus the outcome residual can be \emph{unintendedly} stabilized and lowered. As a result, PS model misspecification does not always inflate $\Delta_{dr}$ because the increase in PS residuals might be offset by the decrease in outcome residuals. An example is in Figure~\ref{fig:figure2}: under poor overlap, misspecification in the propensity model does not increase (in fact, even decreases) the finite-sample error in regions with PS close to 0 and 1, echoing the corresponding deflated outcome residual in Figure~\ref{fig:figure1}.

\section{Application}

\begin{figure}[!ht]
\centering
\includegraphics[width=120mm]{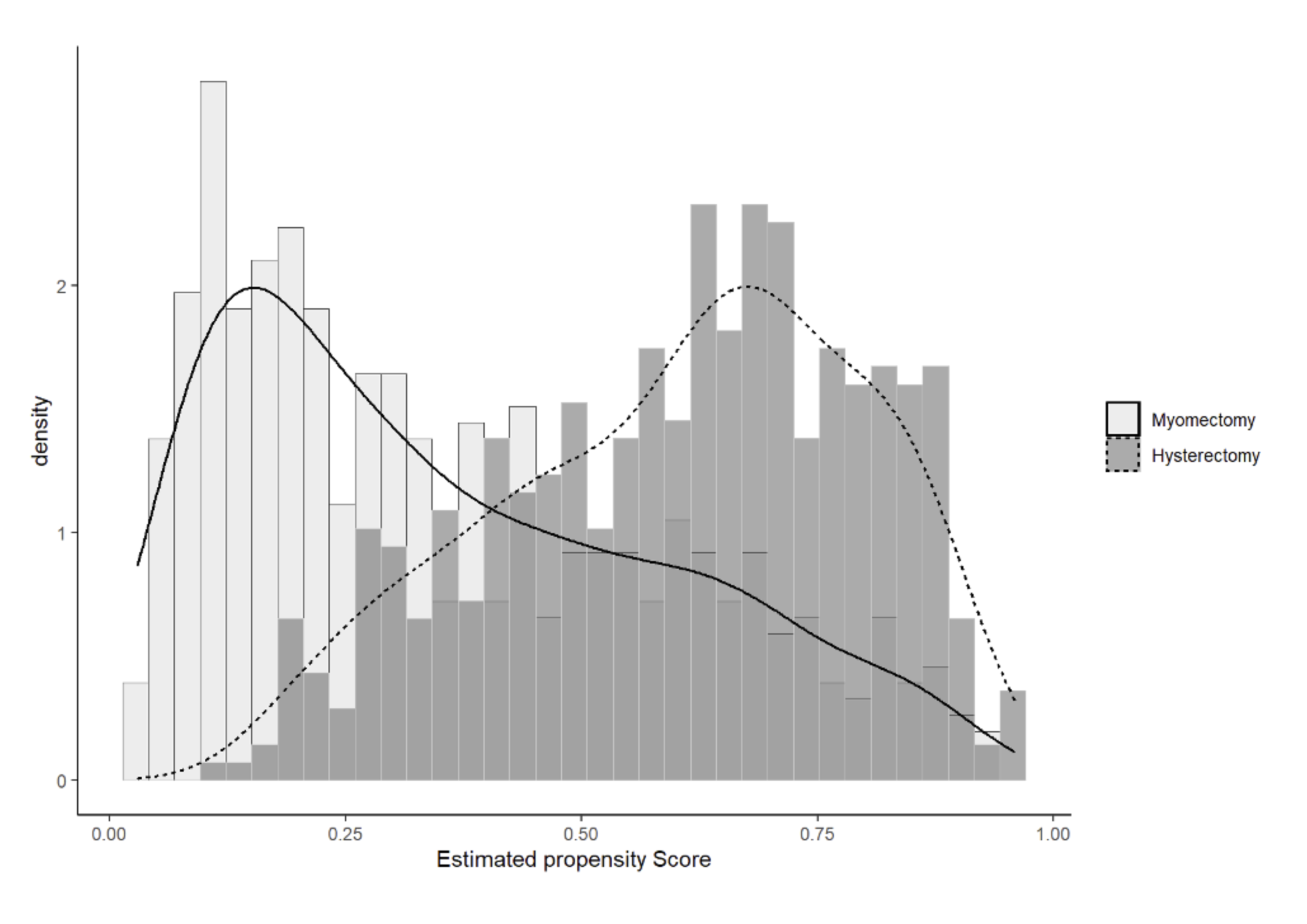}
\caption{Distribution of estimated propensity scores by treatment for uterine fibroids.}
\label{fig:figure3}
\end{figure}

We analyzed the COMPARE-UF Fibroid Registry data to compare the effect on quality of life between hysterectomy and myomectomy for treating uterine fibroids \cite{Stewart2018CompareUF, nicholson2019leiomyomas, laughlin2020CompareUF}. Our dataset contains 1060 patients, with 557 receiving myomectomy and 503 receiving hysterectomy. We studied the endpoint of the UFS total quality of life (UFS-QoL), and identified 16 confounders, including demographics, medical history, baseline QoL components, and clinical features. We specify a logistic model for the PS of receiving hysterectomy and a linear model for the outcome modeling, with main effects of each of the 16 confounders. As shown in the histogram of the estimated propensity scores in Figure~\ref{fig:figure3}, the patients between the two treatments are vastly different, with a sizable number of patients having PS close to 0 or 1.

\begin{table}[!ht]
\centering
\caption{Estimated mean difference in UFS-QoL by different methods.}
\begin{tabular}{lc}
\midrule
\multirow{2}{*}{Estimator} & Treatment effect in UFS-QoL (95\% CI) (SD) \\
 & Hysterectomy $-$ Myomectomy \\ \midrule
Doubly-robust &  \\
\quad No trimming & 4.93 (1.41, 8.45) (1.79) \\
\quad $\delta = 0.05$ & 4.95 (1.45, 8.44) (1.78) \\
\quad $\delta = 0.1$ & 3.73 (0.41, 7.04) (1.69) \\
IPW &  \\
\quad No trimming & 4.57 (0.96, 8.17) (1.84) \\
\quad $\delta = 0.05$ & 4.77 (1.16, 8.38) (1.84) \\
\quad $\delta = 0.1$ & 3.54 (0.24, 6.84) (1.69) \\
Outcome regression & 3.84 (0.54, 7.13) (1.68) \\
Overlap weighting & 3.55 (0.27, 6.83) (1.67) \\ \midrule
\end{tabular}
\label{tab:Compare-UF}
\end{table}

Table~\ref{tab:Compare-UF} displays the estimated causal effects in QoL between hysterectomy and myomectomy by different methods. The results are considerably different: inclusion of units in the region with little overlap, i.e., methods with non-trimmed weights or trimming at $\delta = 0.05$, yields larger effects but also larger standard errors. In contrast, methods that focus on the subpopulation with more overlap, i.e., overlap weighting and trimming at $\delta = 0.1$, yield milder effects with smaller standard errors and agree with the outcome regression estimator. These results suggest potential sizable heterogeneity in treatment effects across subpopulations and necessitate careful consideration in choosing the target population for inference.

\section{Discussion}

We investigate doubly-robust methods for causal inference in observational studies. First, specification of the outcome model has a stronger influence on the DR estimates than specification of the propensity score model, and the dominance increases as the degree of overlap decreases. Second, with poor overlap, the DR estimator generally incurs large variance, and if the outcome model is misspecified, large bias; in fact, the DR estimator can be universally worse than an estimator based solely on a correct outcome model. We provide analytical and empirical explanations of these phenomena.

We emphasize that the DR estimation is not a magic relief to poor overlap, and in fact, it not only inherits but also exacerbates the adverse consequences of lack of overlap by amplifying the prediction error from the outcome model with extreme inverse propensity weights. Therefore, as revealed in our simulations, DR with misspecified outcome model is worse than a correct IPW estimator. This exacting demand on the outcome model largely defeats the purpose DR under poor overlap: augmenting IPW by outcome modeling (or vice versa) only exacerbates bias and variance unless the outcome model is correct, but on the other hand, there is no benefit of using DR if a correct outcome model is already known and available. Therefore, when there is a severe lack of overlap in the study, we recommend analysts to move the goal post to a target population with adequate overlap, via, e.g., matching, overlap weights, or weighting estimators with trimmed weights. When the change in target population is undesirable, one could adopt IPW with calibrated weights, e.g., the stabilized balancing weights \cite{Zubizarreta2015Stableweights} or \cite{Imai2013CBPS} CBPS, to mitigate the extreme weights. In some cases, outcome modeling may also be preferable in terms of both bias and variance, and one may shift to flexible outcome methods, such as machine learning and Bayesian nonparametric models \cite{Hill2011BayesCausal}.

\bibliographystyle{apalike}
\bibliography{DR}

\newpage
\begin{center}
    \Large\bf Supplementary Materials
\end{center}

This supplementary material accompanies the main manuscript and is organized into two sections. Section~A provides additional tables and figures for the primary simulation results presented in the main text. Section~B presents figures from secondary simulations examining estimator performance across a range of sample sizes, each based on 10,000 replicated datasets.

\section*{Section A: Additional results for primary simulation}

\subsection*{Tables}
\vspace{0.2em}
\begin{itemize}
    \item[A1.] Proportion of propensity scores outside the range of $[0.05, 0.95]$
    \item[A2.] RMSE and bias under misspecification by omitting variables
    \item[A3.] Mean and MAV of finite-sample error under misspecification by omitting variables
    \item[A4.] Scaled relative MAV for intervals not shown in Figure~1 panel~(b)
    \item[A5.] Scaled relative MAV for intervals not shown in Figure~2 panel~(b)
    \item[A6.] Scaled relative MAV for intervals not shown in Figure~A1 panel~(b)
    \item[A7.] Scaled relative MAV for intervals not shown in Figure~A2 panel~(b)
\end{itemize}

\subsection*{Figures}
\vspace{0.2em}
\begin{itemize}
    \item[A1.] Scaled relative MAV of $R_i^e$ and $R_i^y$ under misspecification by omitting variables
    \item[A2.] Scaled relative MAV of $R_i^e R_i^y$ under misspecification by omitting variables
    \item[A3.] Scaled relative MAV of $R_i^e$ and $R_i^y$ for panels~(c) and~(d) not shown in Figures~1 and~A1
    \item[A4.] Scaled relative MAV of $R_i^e R_i^y$ for panels~(c) and~(d) not shown in Figures~2 and~A2
    \item[A5.] Distributions of true propensity scores by treatment group
    \item[A6.] Distributions of true propensity scores
\end{itemize}

\vspace{0.5em}

\section*{Section B: Secondary simulation results across sample sizes}

\subsection*{Figures}
\vspace{0.2em}
\begin{itemize}
    \item[B1.] Absolute bias for different estimators across sample sizes
    \item[B2.] RMSE for different estimators across sample sizes
    \item[B3.] Absolute bias for DR estimators under misspecification by wrong functional form
    \item[B4.] RMSE for DR estimators under misspecification by wrong functional form
    \item[B5.] Absolute bias for DR estimators under misspecification by omitting variables
    \item[B6.] RMSE for DR estimators under misspecification by omitting variables
\end{itemize}

\clearpage
\FloatBarrier

\setcounter{section}{0}
\renewcommand{\thesection}{\Alph{section}}
\setcounter{equation}{0}
\numberwithin{equation}{section}

\setcounter{table}{0}
\setcounter{figure}{0}
\renewcommand{\thetable}{A\arabic{table}}
\renewcommand{\thefigure}{A\arabic{figure}}

\section{Additional results for primary simulation}

\begin{table}[!ht]
\centering
\caption{Proportion of propensity scores outside the range of $(0.05, 0.95)$ for various simulation settings.}
\begin{tabular}{cccc}
\midrule
 Treatment prevalence & $d$ & $\alpha_0$ & Proportion (\%) of PS outside of $(0.05, 0.95)$ \\ \midrule
\multirow{2}{*}{0.4} & 1 & $-$0.05 & 0 \\
 & 3 & 0.37 & 36 \\ \midrule
\multirow{2}{*}{0.1} & 1 & $-$2.13 & 36 \\
 & 3 & $-$3.16 & 68 \\ \midrule
\end{tabular}

\noindent{\footnotesize Values are simulated based on a large sample size of 100,000.}
\label{tab:A1}
\end{table}

\FloatBarrier

\begin{table}[!ht]
\caption{RMSE and bias for estimators with one of the propensity and outcome models being misspecified, or both models being misspecified. The misspecification is incurred by omitting variables (omitting $X_3$ and $X_4$).}
\resizebox{\textwidth}{!}{
\begin{tabular}{lllcccccccc}
\midrule
 \multicolumn{2}{c}{\multirow{3}{*}{Estimator}} & \multirow{3}{*}{\shortstack{Misspecified\\model}} & \multicolumn{4}{c}{Treatment prevalence = 0.4} & \multicolumn{4}{c}{Treatment prevalence = 0.1} \\
 &  &  & \multicolumn{2}{c}{$d = 1$} & \multicolumn{2}{c}{$d = 3$} & \multicolumn{2}{c}{$d = 1$} & \multicolumn{2}{c}{$d = 3$} \\ \cmidrule{4-11}
 &  &  & RMSE & Bias & RMSE & Bias & RMSE & Bias & RMSE & Bias \\ \midrule
\multirow{6}{*}{\shortstack{Doubly-\\robust}} & \multirow{3}{*}{No trimming} & PS & 0.11 & 0.00 & 0.22 & 0.00 & 0.23 & 0.00 & 0.95 & $-$0.01 \\
 &  & Outcome & 0.12 & 0.00 & 1.17 & $-$0.02 & 0.34 & $-$0.03 & 4.85 & $-$0.36 \\
 &  & Both & 0.39 & $-$0.36 & 1.05 & $-$1.00 & 0.49 & $-$0.37 & 1.80 & $-$1.14 \\ \cmidrule{2-11}
 & \multirow{3}{*}{$\delta = 0.05$} & PS & 0.11 & 0.00 & 0.15 & 0.00 & 0.20 & 0.00 & 0.27 & $-$0.01 \\
 &  & Outcome & 0.11 & 0.00 & 0.18 & $-$0.01 & 0.22 & $-$0.02 & 0.28 & $-$0.04 \\
 &  & Both & 0.39 & $-$0.36 & 1.00 & $-$0.98 & 0.46 & $-$0.37 & 1.00 & $-$0.95 \\ \midrule
\multirow{2}{*}{IPW} & No trimming & PS & 0.41 & $-$0.37 & 1.40 & $-$1.28 & 0.70 & $-$0.41 & 2.42 & $-$2.14 \\
 & $\delta = 0.05$ & PS & 0.41 & $-$0.36 & 1.03 & $-$1.00 & 0.48 & $-$0.37 & 1.02 & $-$0.96 \\ \midrule
\multicolumn{2}{l}{Outcome regression} & Outcome & 0.39 & $-$0.36 & 0.95 & $-$0.93 & 0.47 & $-$0.37 & 1.34 & $-$1.20 \\ \midrule
\multicolumn{2}{l}{Overlap weighting} & PS & 0.39 & $-$0.36 & 0.98 & $-$0.97 & 0.43 & $-$0.36 & 0.96 & $-$0.93 \\ \midrule
\end{tabular}
}
\noindent{\footnotesize Results from trimming with $\delta = 0.1$ are similar to that with $\delta = 0.05$ and are thus not displayed.}
\label{tab:A2}
\end{table}

\FloatBarrier

\begin{table}[!ht]
\caption{Mean and mean absolute value (MAV) of finite-sample error in different (sub)populations with one of the propensity and outcome models being misspecified by omitting variables (omitting $X_3$ and $X_4$).}
\resizebox{\textwidth}{!}{
\begin{tabular}{lllcccccccc}
\midrule
 \multicolumn{2}{c}{\multirow{3}{*}{Model specification}} & \multirow{3}{*}{Population} & \multicolumn{4}{c}{Treatment prevalence = 0.4} & \multicolumn{4}{c}{Treatment prevalence = 0.1} \\
 &  &  & \multicolumn{2}{c}{$d = 1$} & \multicolumn{2}{c}{$d = 3$} & \multicolumn{2}{c}{$d = 1$} & \multicolumn{2}{c}{$d = 3$} \\ \cmidrule{4-11}
 &  &  & Mean & MAV & Mean & MAV & Mean & MAV & Mean & MAV \\ \midrule
\multirow{6}{*}{\shortstack{Misspecified\\model}} & \multirow{3}{*}{PS model} & Overall & 0.00 & 0.07 & 0.00 & 0.15 & 0.00 & 0.17 & $-$0.01 & 0.46 \\
 &  & Tail & 0.00 & 0.29 & $-$0.02 & 0.58 & 0.00 & 0.47 & $-$0.02 & 0.64 \\
 &  & Bulk & 0.00 & 0.08 & 0.00 & 0.11 & 0.00 & 0.20 & $-$0.01 & 0.46 \\ \cmidrule{2-11}
 & \multirow{3}{*}{\shortstack{Outcome\\model}} & Overall & 0.00 & 0.08 & $-$0.02 & 0.45 & $-$0.03 & 0.24 & $-$0.36 & 1.17 \\
 &  & Tail & $-$0.03 & 0.59 & $-$0.12 & 2.96 & $-$0.09 & 1.10 & $-$0.26 & 2.59 \\
 &  & Bulk & 0.00 & 0.11 & $-$0.01 & 0.16 & $-$0.02 & 0.27 & $-$0.29 & 1.01 \\ \midrule
\end{tabular}
}
\label{tab:A3}
\end{table}

\FloatBarrier

\begin{table}[!ht]
\centering
\caption{Scaled relative MAV for intervals not shown in Figure~\ref{fig:figure1} panel (b).}
\begin{tabular}{cccc}
\midrule
PS interval & Residual type & Misspecified model & Scaled relative MAV \\ \midrule
\multirow{3}{*}{$(0.00, 0.01]$} & \multirow{2}{*}{Outcome} & Outcome & 1.43 \\
 &  & PS & 0.01 \\ \cmidrule{2-4}
 & PS & PS & 0.38 \\ \midrule
\multirow{3}{*}{$(0.01, 0.02]$} & \multirow{2}{*}{Outcome} & Outcome & 0.37 \\
 &  & PS & 0.04 \\ \cmidrule{2-4}
 & PS & PS & 0.11 \\ \midrule
\multirow{3}{*}{$(0.02, 0.03]$} & \multirow{2}{*}{Outcome} & Outcome & 0.19 \\
 &  & PS & 0.03 \\ \cmidrule{2-4}
 & PS & PS & 0.06 \\ \midrule
\multirow{3}{*}{$(0.98, 0.99]$} & \multirow{2}{*}{Outcome} & Outcome & 0.20 \\
 &  & PS & 0.01 \\ \cmidrule{2-4}
 & PS & PS & 0.04 \\ \midrule
\multirow{3}{*}{$(0.99, 1.00)$} & \multirow{2}{*}{Outcome} & Outcome & 0.33 \\
 &  & PS & 0.01 \\ \cmidrule{2-4}
 & PS & PS & 0.05 \\ \midrule
\end{tabular}
\label{tab:A4}
\end{table}

\FloatBarrier

\begin{table}[!ht]
\centering
\caption{Scaled relative MAV for intervals not shown in Figure~\ref{fig:figure2} panel (b).}
\begin{tabular}{ccc}
\midrule
\multirow{2}{*}{PS interval} & \multicolumn{2}{c}{Misspecified model} \\ \cmidrule{2-3}
 & PS model & Outcome model \\ \midrule
$(0.00, 0.01]$ & 0.05 & 0.53 \\
$(0.01, 0.02]$ & 0.04 & 0.16 \\
$(0.02, 0.03]$ & 0.03 & 0.09 \\
$(0.98, 0.99]$ & 0.01 & 0.10 \\
$(0.99, 1.00)$ & 0.01 & 0.18 \\ \midrule
\end{tabular}
\label{tab:A5}
\end{table}

\FloatBarrier

\begin{table}[!ht]
\centering
\caption{Scaled relative MAV for intervals not shown in Figure~\ref{fig:A1} panel (b).}
\begin{tabular}{cccc}
\midrule
PS interval & Residual type & Misspecified model & Scaled relative MAV \\ \midrule
\multirow{3}{*}{$(0.00, 0.01]$} & \multirow{2}{*}{Outcome} & Outcome & 0.63 \\
 &  & PS & 0.01 \\ \cmidrule{2-4}
 & PS & PS & 0.29 \\ \midrule
\multirow{3}{*}{$(0.01, 0.02]$} & \multirow{2}{*}{Outcome} & Outcome & 0.31 \\
 &  & PS & 0.03 \\ \cmidrule{2-4}
 & PS & PS & 0.12 \\ \midrule
\multirow{3}{*}{$(0.02, 0.03]$} & \multirow{2}{*}{Outcome} & Outcome & 0.18 \\
 &  & PS & 0.03 \\ \cmidrule{2-4}
 & PS & PS & 0.07 \\ \midrule
\multirow{3}{*}{$(0.98, 0.99]$} & \multirow{2}{*}{Outcome} & Outcome & 0.12 \\
 &  & PS & 0.01 \\ \cmidrule{2-4}
 & PS & PS & 0.05 \\ \midrule
\multirow{3}{*}{$(0.99, 1.00)$} & \multirow{2}{*}{Outcome} & Outcome & 0.13 \\
 &  & PS & 0.00 \\ \cmidrule{2-4}
 & PS & PS & 0.08 \\ \midrule
\end{tabular}
\label{tab:A6}
\end{table}

\begin{table}[!ht]
\centering
\caption{Scaled relative MAV for intervals not shown in Figure~\ref{fig:A2} panel (b).}
\begin{tabular}{ccc}
\midrule
\multirow{2}{*}{PS interval} & \multicolumn{2}{c}{Misspecified model} \\ \cmidrule{2-3}
 & PS model & Outcome model \\ \midrule
$(0.00, 0.01]$ & 0.05 & 0.36 \\
$(0.01, 0.02]$ & 0.03 & 0.14 \\
$(0.02, 0.03]$ & 0.03 & 0.09 \\
$(0.98, 0.99]$ & 0.01 & 0.06 \\
$(0.99, 1.00)$ & 0.01 & 0.09 \\ \midrule
\end{tabular}
\label{tab:A7}
\end{table}

\FloatBarrier

\begin{figure}[!ht]
\centering
\includegraphics[width=\textwidth]{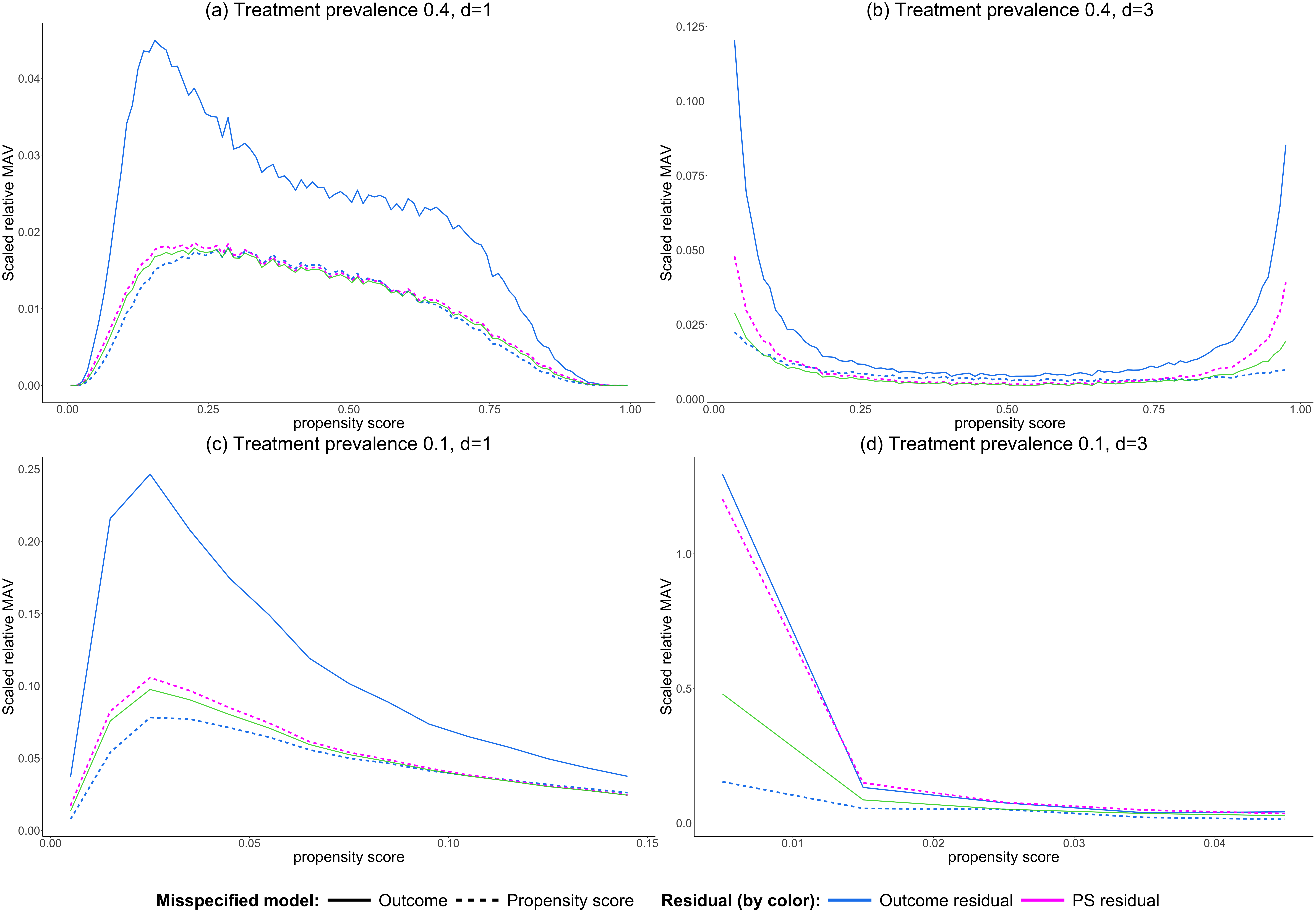}
\caption{Scaled relative MAV of $R_i^e$ and $R_i^y$. Misspecification is incurred by omitting variables (omitting $X_3$ and $X_4$). The green line ($\Phi(l)$) indicates no change in relative scale. Large values at tails of panel (b) and the rest of panels (c) and (d) are delegated to Table~\ref{tab:A6} and Figure~\ref{fig:A3}.}
\label{fig:A1}
\end{figure}

\FloatBarrier

\begin{figure}[!ht]
\centering
\includegraphics[width=\textwidth]{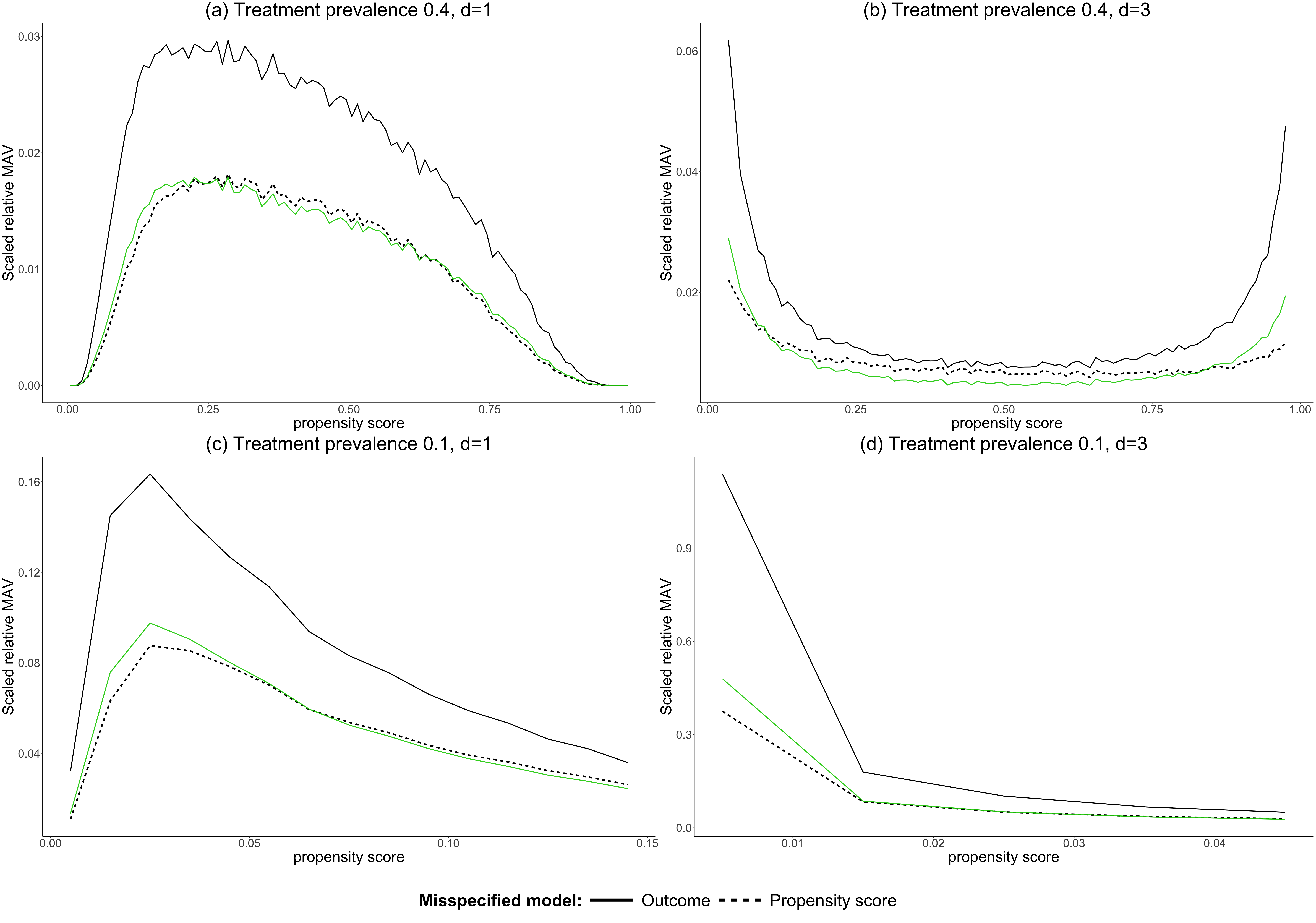}
\caption{Scaled relative MAV of $R_i^e R_i^y$. Misspecification is incurred by omitting variables (omitting $X_3$ and $X_4$). The green line ($\Phi(l)$) indicates no change in relative scale. Large values at tails of panel (b) and the rest of panels (c) and (d) are delegated to Table~\ref{tab:A7} and Figure~\ref{fig:A4}.}
\label{fig:A2}
\end{figure}

\FloatBarrier

\begin{figure}[!ht]
\centering
\includegraphics[width=\textwidth]{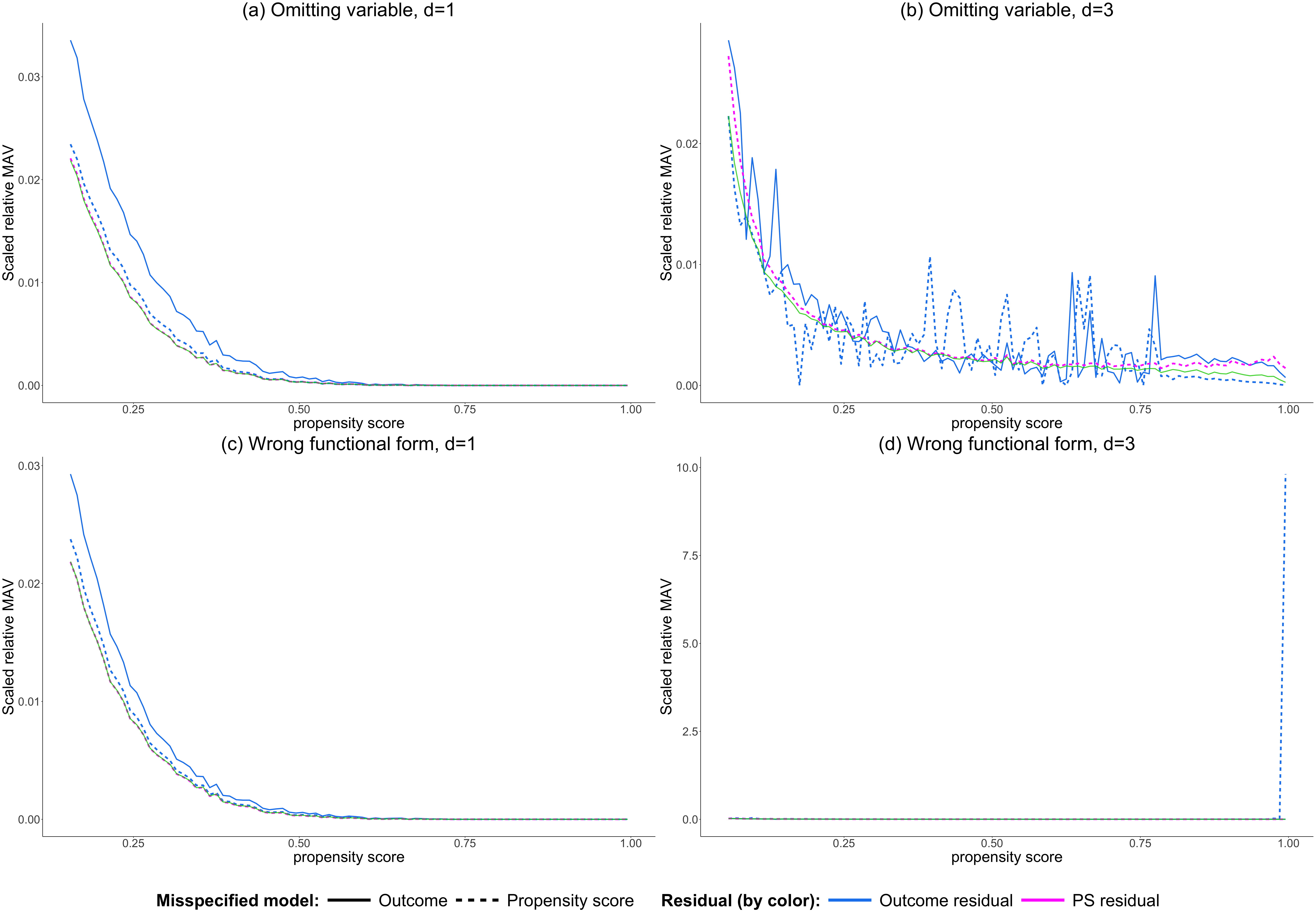}
\caption{Scaled relative MAV of $R_i^e$ and $R_i^y$ for the rest of panels (c) and (d) not shown in Figure~\ref{fig:figure1} and Figure~\ref{fig:A1}. $d=1$ corresponds to panels (c) of Figures~\ref{fig:figure1} and \ref{fig:A1}, and $d=3$ corresponds to panels (d) of Figures~\ref{fig:figure1} and \ref{fig:A1}.}
\label{fig:A3}
\end{figure}

\FloatBarrier

\begin{figure}[!ht]
\centering
\includegraphics[width=\textwidth]{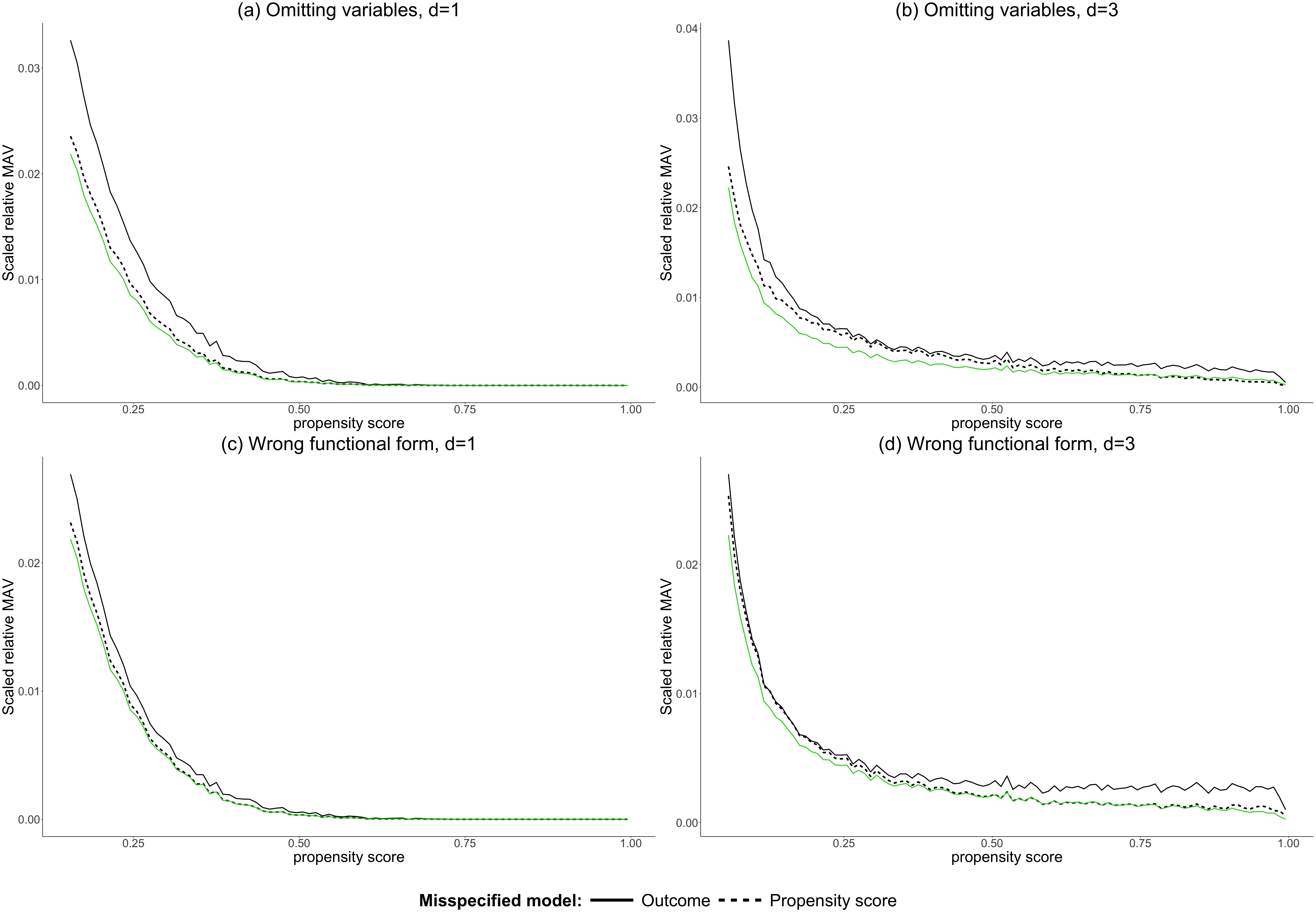}
\caption{Scaled relative MAV of $R_i^e R_i^y$ for the rest of panels (c) and (d) not shown in Figure~\ref{fig:figure2} and Figure~\ref{fig:A2}. $d=1$ corresponds to panels (c) of Figures~\ref{fig:figure2} and \ref{fig:A2}, and $d=3$ corresponds to panels (d) of Figures~\ref{fig:figure2} and \ref{fig:A2}.}
\label{fig:A4}
\end{figure}

\FloatBarrier

\begin{figure}[!ht]
\centering
\includegraphics[width=\textwidth]{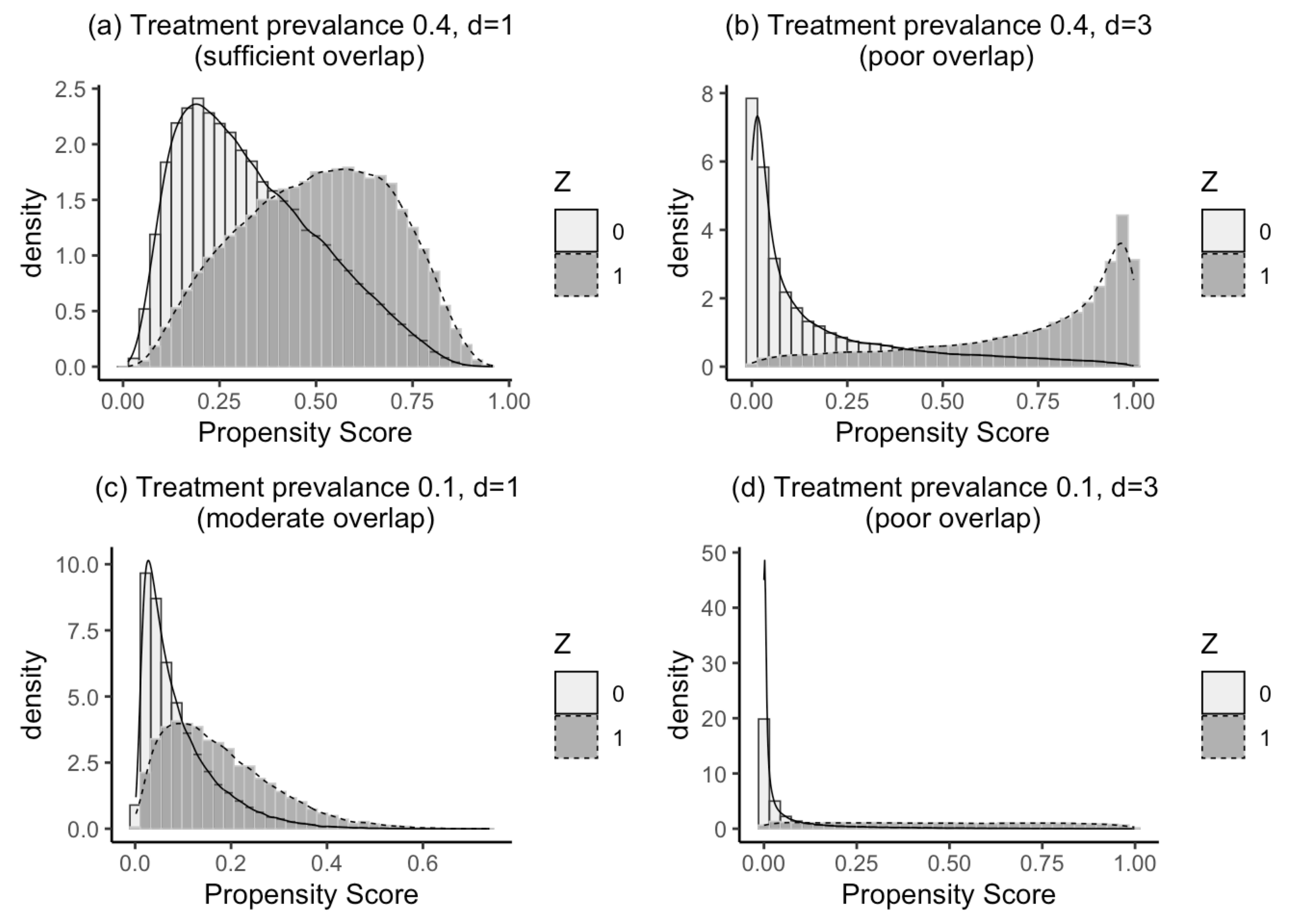}
\caption{Distributions of true propensity scores by treatment groups (evaluated by large-sample dataset with size of 100,000).}
\label{fig:A5}
\end{figure}

\FloatBarrier

\begin{figure}[!ht]
\centering
\includegraphics[width=\textwidth]{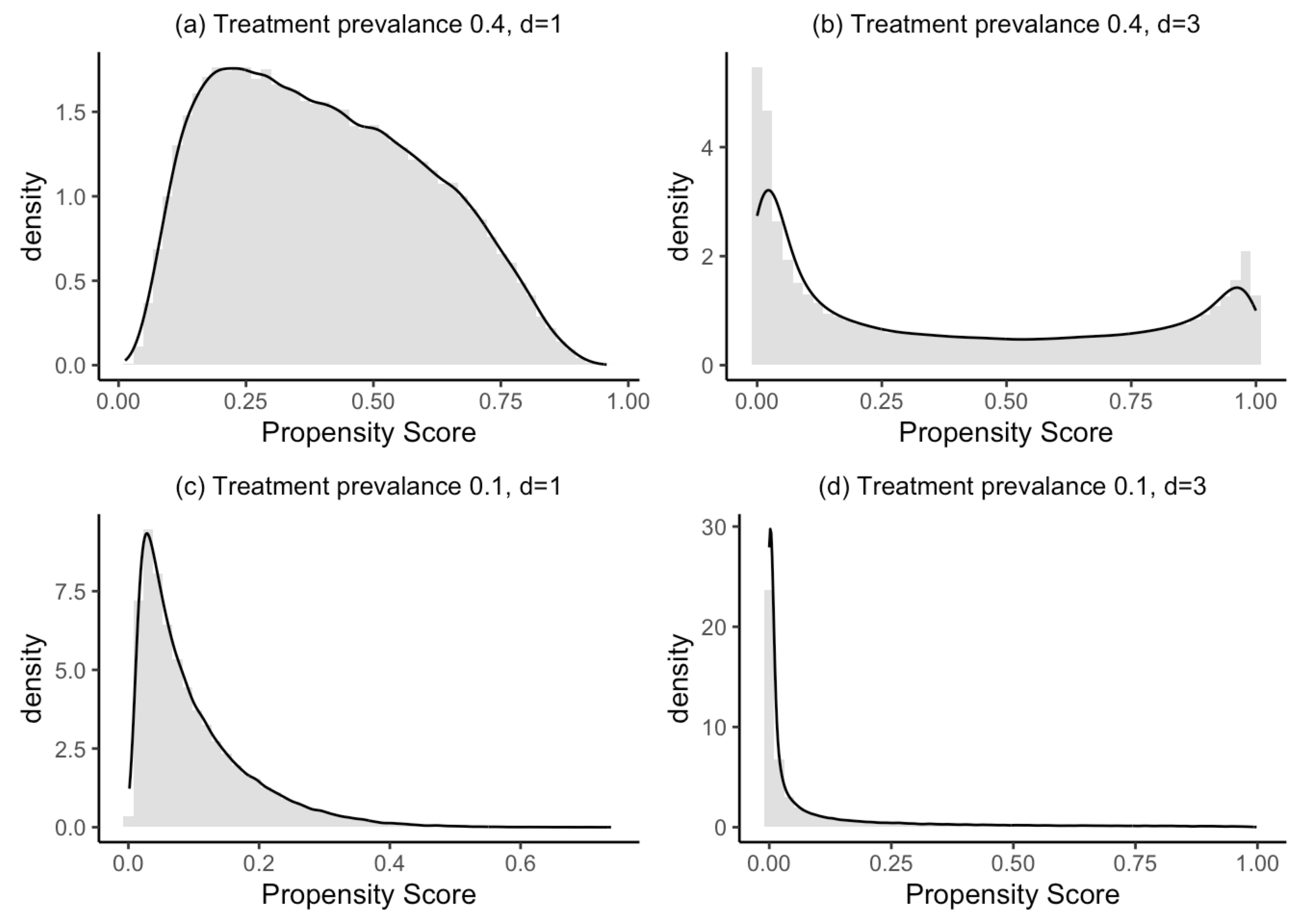}
\caption{Distributions of true propensity scores (evaluated by large-sample dataset with size of 100,000).}
\label{fig:A6}
\end{figure}

\clearpage
\FloatBarrier

\setcounter{table}{0}
\setcounter{figure}{0}
\renewcommand{\thetable}{B\arabic{table}}
\renewcommand{\thefigure}{B\arabic{figure}}

\section{Secondary simulation results across sample sizes}

\begin{figure}[!ht]
\centering
\includegraphics[width=\textwidth]{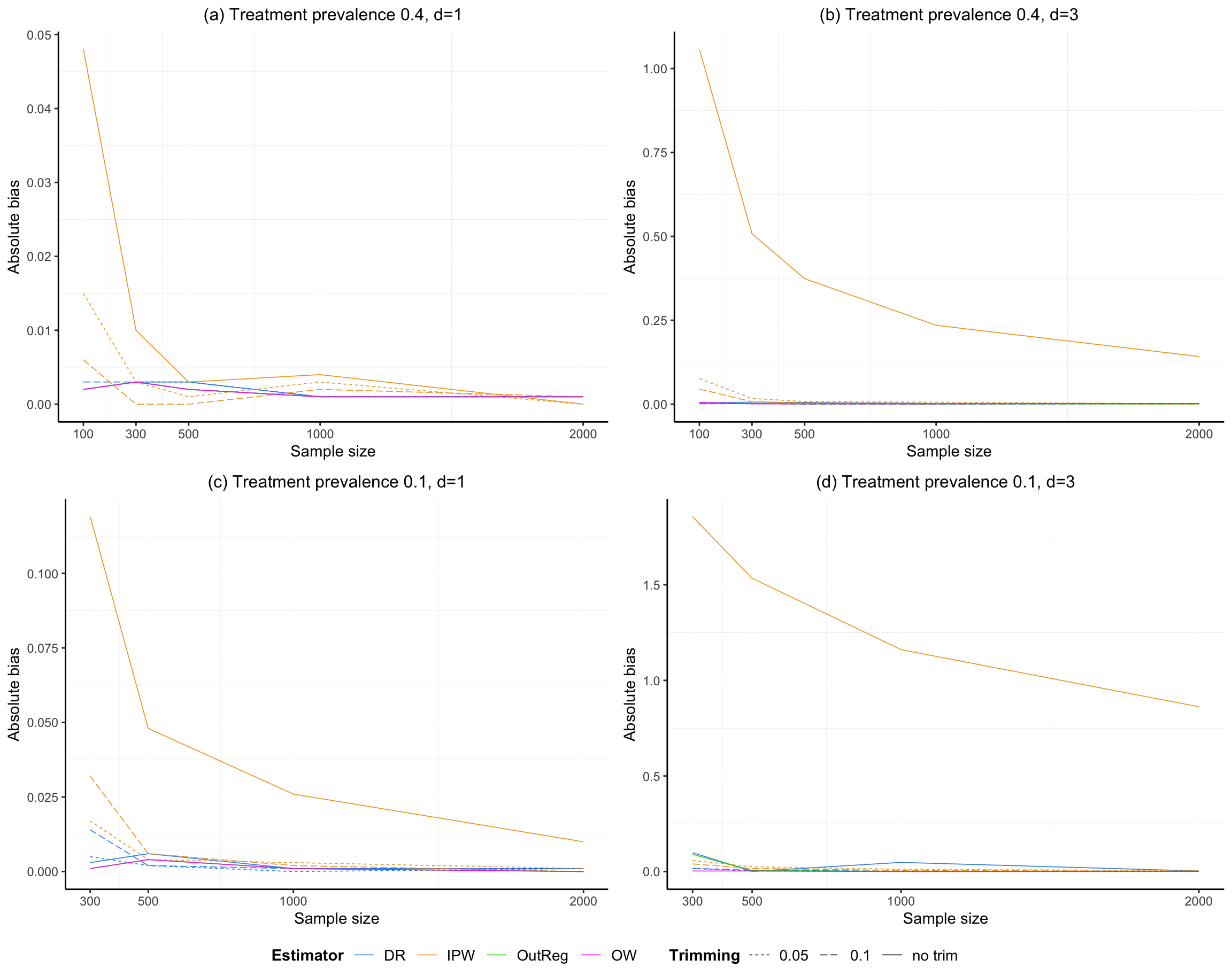}
\caption{Absolute bias for different estimators under each sample size.}
\label{fig:B1}
\end{figure}

\FloatBarrier

\begin{figure}[!ht]
\centering
\includegraphics[width=\textwidth]{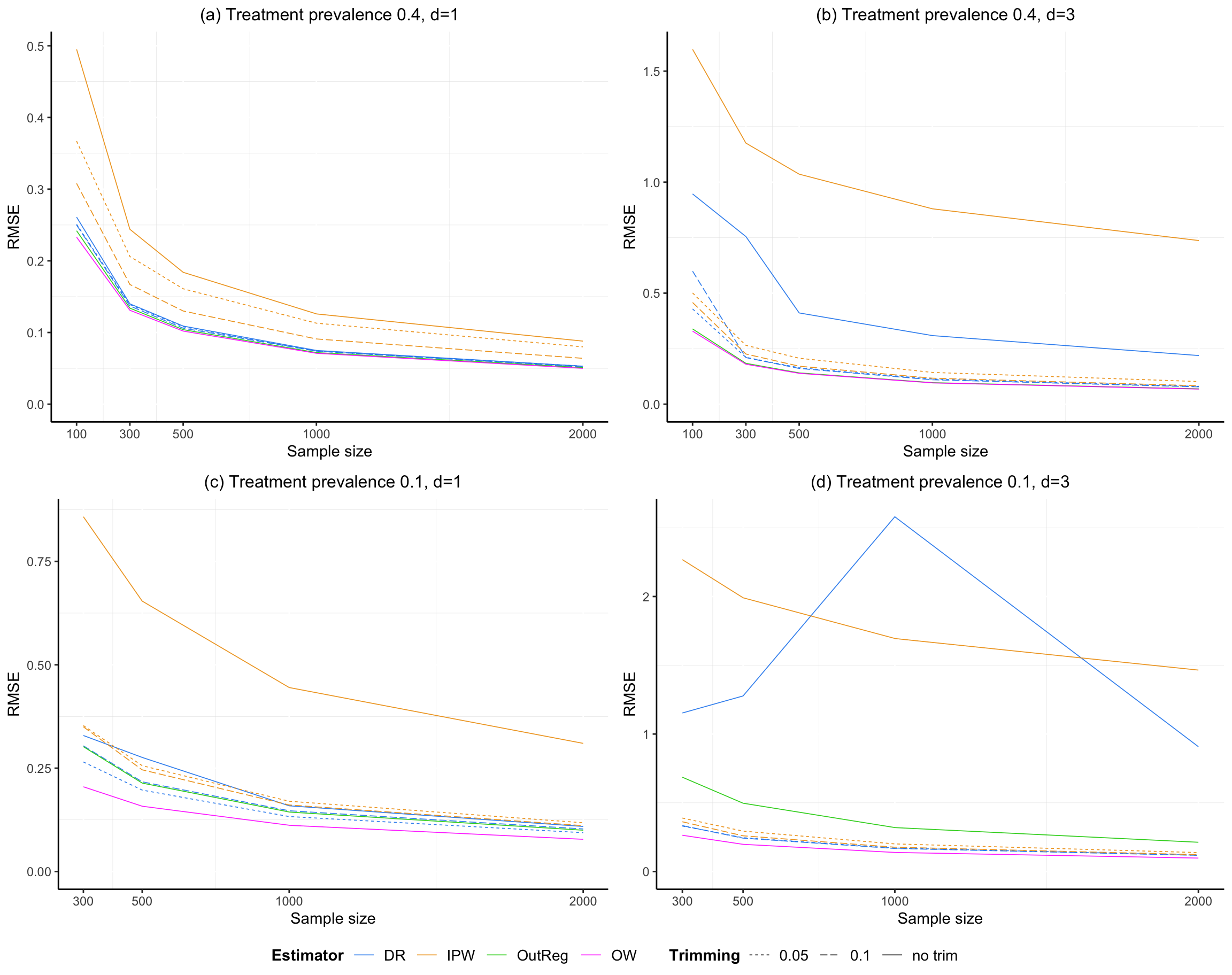}
\caption{RMSE (root mean squared error) for different estimators under each sample size.}
\label{fig:B2}
\end{figure}

\FloatBarrier

\begin{figure}[!ht]
\centering
\includegraphics[width=\textwidth]{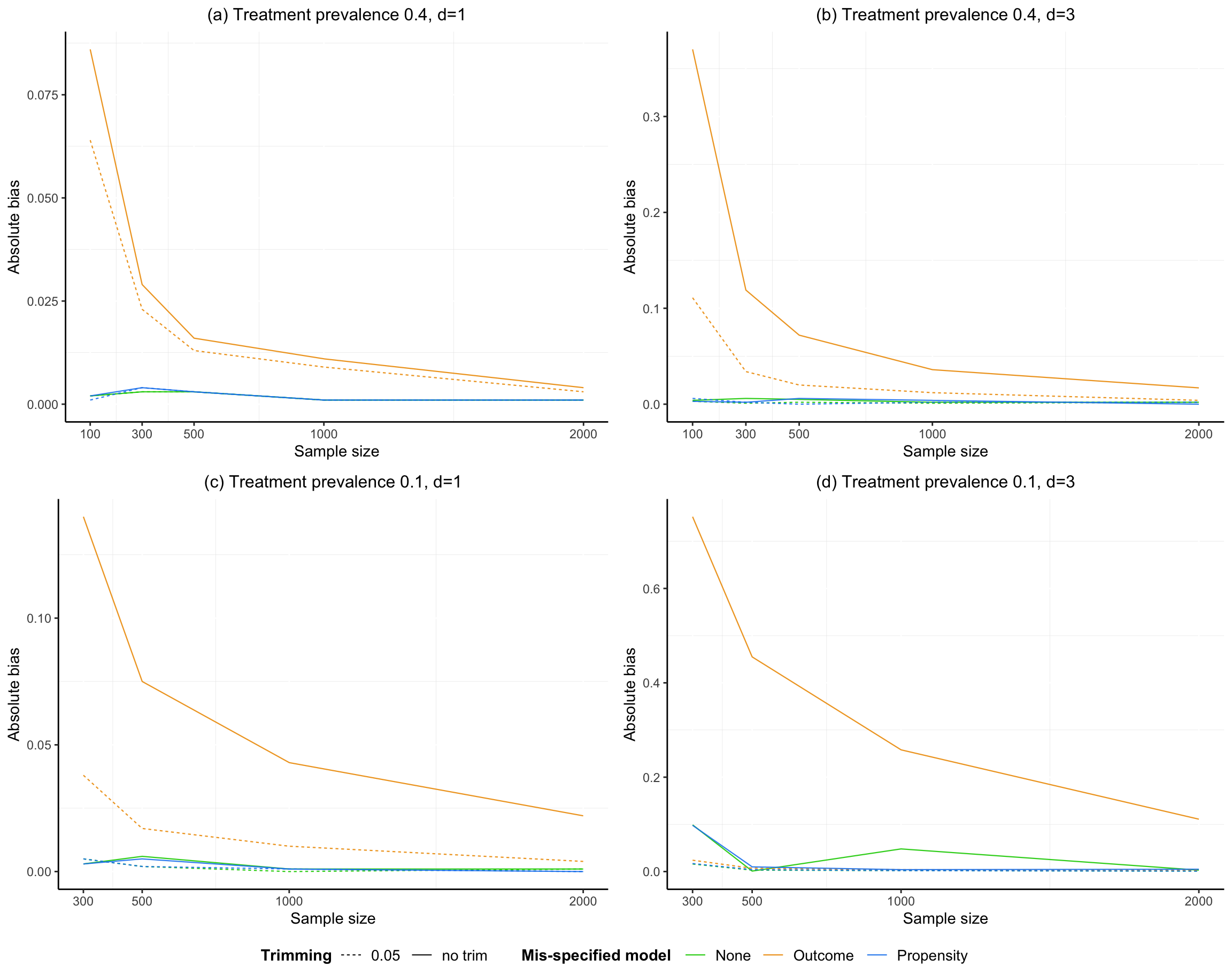}
\caption{Absolute bias for DR estimators (without and with trimming of PS at 0.05) when one of the propensity score or the outcome model is misspecified. Misspecification is incurred by wrong functional form (mistaking $X_3$ by $X_3^2$).}
\label{fig:B3}
\end{figure}

\FloatBarrier

\begin{figure}[!ht]
\centering
\includegraphics[width=\textwidth]{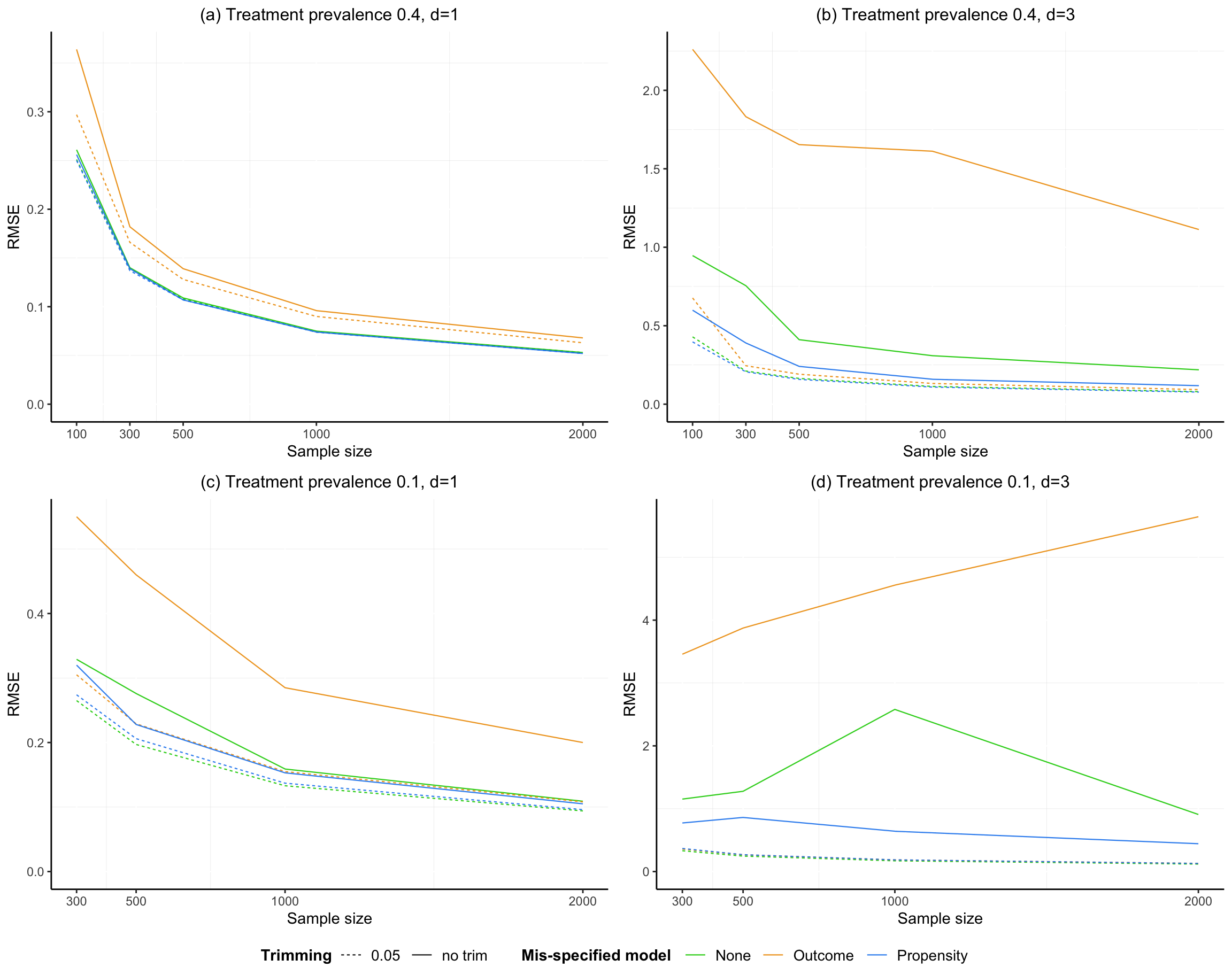}
\caption{RMSE for DR estimators (without and with trimming of PS at 0.05) when one of the propensity score or the outcome model is misspecified. Misspecification is incurred by wrong functional form (mistaking $X_3$ by $X_3^2$).}
\label{fig:B4}
\end{figure}

\FloatBarrier

\begin{figure}[!ht]
\centering
\includegraphics[width=\textwidth]{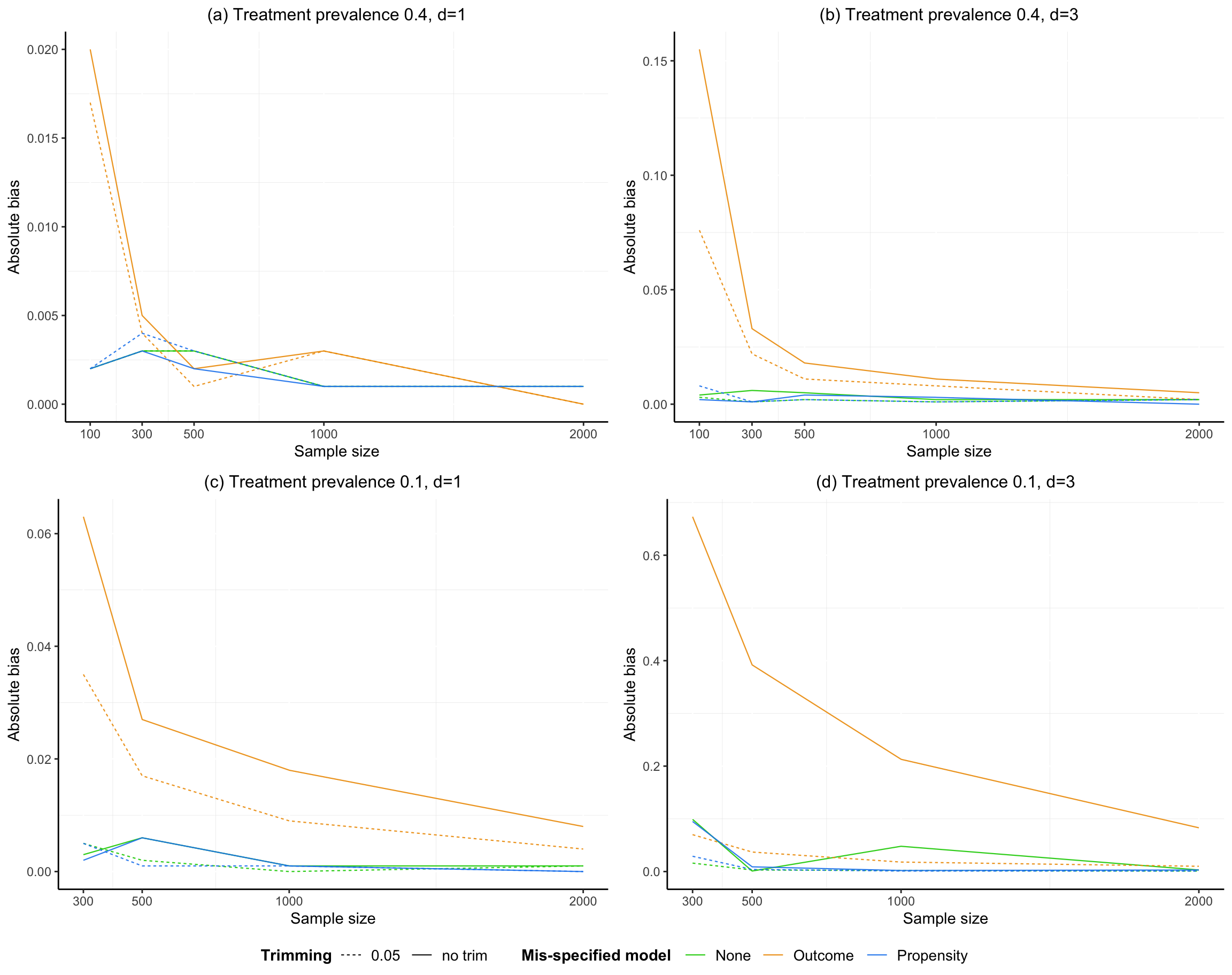}
\caption{Absolute bias for DR estimators (without and with trimming of PS at 0.05) when one of the propensity score or the outcome model is misspecified. Misspecification is incurred by omitting variables (omitting $X_3$ and $X_4$).}
\label{fig:B5}
\end{figure}

\FloatBarrier

\begin{figure}[!ht]
\centering
\includegraphics[width=\textwidth]{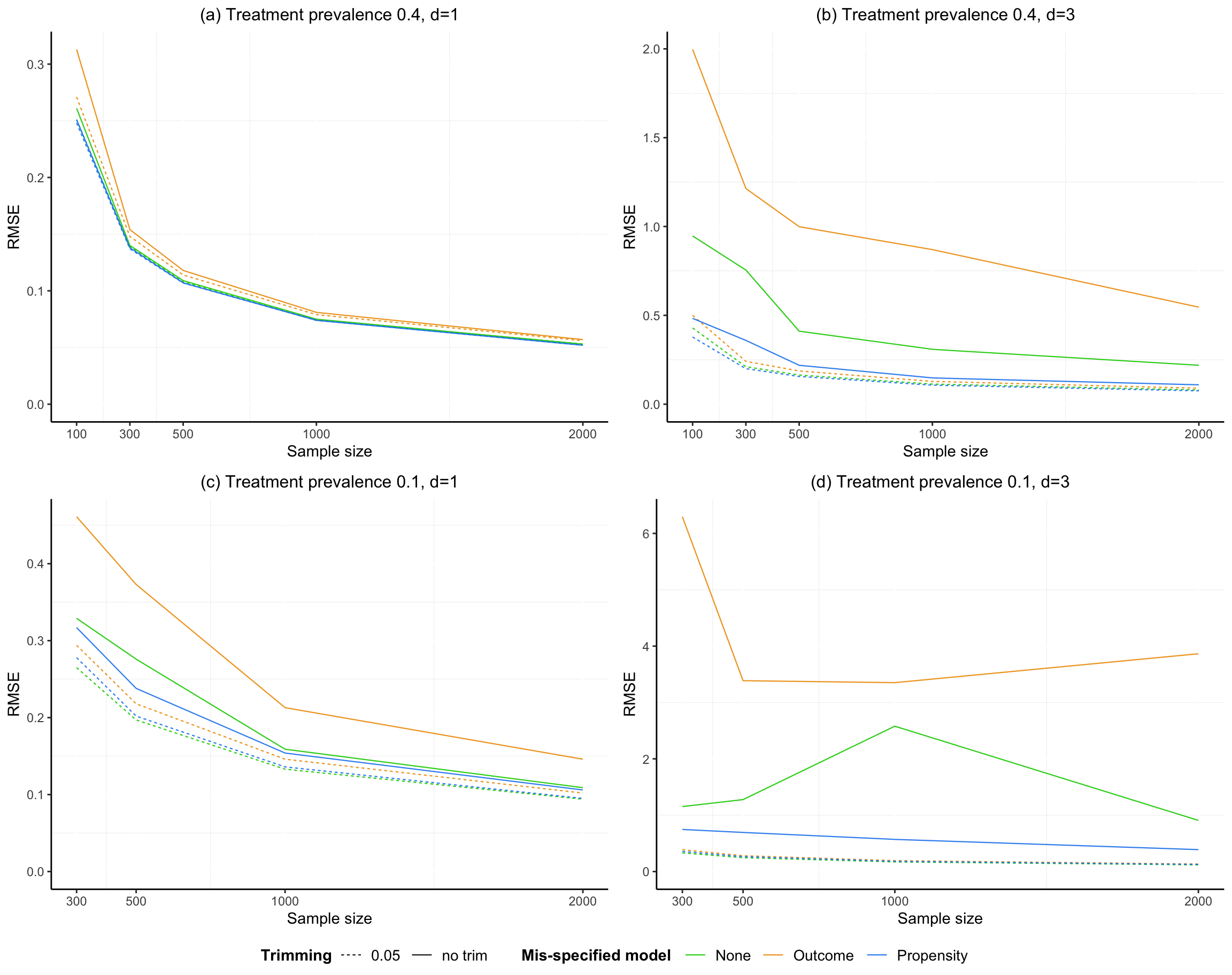}
\caption{RMSE for DR estimators (without and with trimming of PS at 0.05) when one of the propensity score or the outcome model is misspecified. Misspecification is incurred by omitting variables (omitting $X_3$ and $X_4$).}
\label{fig:B6}
\end{figure}

\end{document}